\documentclass[aip, reprint, amsmath, amssymb, superscriptaddress,floatfix,twocolumn]{revtex4-1}
\usepackage{amsmath}
\usepackage{float}

\newcommand{\Sec}[1]{Sec.\,\ref{#1}}

\newcommand{\be}{\begin{equation}}
	\newcommand{\ee}{\end{equation}}
\newcommand{\bea}{\begin{eqnarray}}
	\newcommand{\eea}{\end{eqnarray}}
\newcommand{\Fig}[1]{Fig.\,\ref{#1}}
\newcommand{\Eq}[1]{Eq.\,(\ref{#1})}
\newcommand{\etal}{{\it et al. }}
\newcommand{\la}{\langle}
\newcommand{\ra}{\rangle}

\newcommand{\RNum}[1]{\uppercase\expandafter{\romannumeral #1\relax}}

\usepackage{graphicx}
\usepackage{dcolumn}
\usepackage{bm}
\usepackage{sidecap}
\usepackage{braket}
\usepackage{color}
\usepackage{amssymb}
\usepackage{csquotes}

\begin{document}
\title{Nonequilibrium reaction rate theory: Formulation and implementation within the hierarchical equations of motion approach 
}

\author{Yaling Ke}
\email{yaling.ke@physik.uni-freiburg.de}
\affiliation{
	Institute of Physics, Albert-Ludwig University Freiburg, Hermann-Herder-Strasse 3, 79104 Freiburg, Germany
}

\author{Christoph Kaspar}
\affiliation{
	Institute of Physics, Albert-Ludwig University Freiburg, Hermann-Herder-Strasse 3, 79104 Freiburg, Germany
}

\author{Andr{\'e} Erpenbeck}
\affiliation{Department of Physics, University of Michigan, Ann Arbor, Michigan 48109, USA}
\author{Uri Peskin}
\affiliation{
Schulich Faculty of Chemistry, Technion-Israel Institute of
Technology, Haifa 32000, Israel
}

\author{Michael Thoss}
\email{michael.thoss@physik.uni-freiburg.de}
\affiliation{
Institute of Physics, Albert-Ludwig University Freiburg, Hermann-Herder-Strasse 3, 79104 Freiburg, Germany
}
\affiliation{
EUCOR Centre for Quantum Science and Quantum Computing, Albert-Ludwig
University Freiburg, Hermann-Herder-Strasse 3, 79104 Freiburg, Germany
}

\begin{abstract}
The study of chemical reactions in environments under nonequilibrium conditions has been of interest recently in a variety of contexts, including current-induced reactions in molecular junctions and scanning tunneling microscopy experiments. In this work, we outline a fully quantum mechanical, numerically exact approach to describe chemical reaction rates in such nonequilibrium situations. The approach is based on an extension of the flux correlation function formalism to nonequilibrium conditions and uses a mixed real and imaginary time hierarchical equations of motion approach for the calculation of rate constants. As a specific example, we investigate current-induced intramolecular proton transfer reactions in a molecular junction for different applied bias voltages and molecule-lead coupling strengths. 
\end{abstract}	
\maketitle

\section{Introduction}

Chemical reactions are often characterized by thermal rate constants. The theory of chemical reaction rates is well developed for reactions in the gas phase or in condensed phases, including a variety of approaches that range from transition state theories or approaches that use classical dynamics to numerically exact fully quantum mechanical methods.\cite{wyatt1996dynamics,RevModPhys.62.251,nitzan2006chemical,miller1975semiclassical,miller1975path,chandler1978statistical,truhlar1983current,Topaler_1994_J.Chem.Phys._p7500,geva2001quantum,pollak2005reaction,manthe1995new,manthe2008state,wang2000forward,Richardson_2014_J.Chem.Phys._p74106} A particular efficient approach to calculating thermal reaction rates is the flux correlation function formalism.\cite{RevModPhys.62.251,Miller_J.Chem.Phys._1983_p4889--4898,Thompson_J.Chem.Phys._1997_p142--150,manthe2002reaction,Wang_J.Chem.Phys._2006_p174502,Craig_J.Chem.Phys._2007_p144503}
 A basic assumption of these approaches is that the reaction occurs in equilibrium, i.e. the reactants are in equilibrium at a given temperature.

This assumption is not fulfilled for chemical reactions under nonequilibrium conditions, which have been of interest in different contexts in recent years. Examples include chemical reactions induced in scanning tunneling microscopy (STM) experiments at molecule-metal interfaces \cite{PhysRevLett.100.166101,PhysRevB.79.035423,liljeroth2007current,PhysRevB.81.045402,peller2020sub,ladenthin2015hot,kumagai2014controlling,boeckmann2016direct} as well as current-induced chemical reactions in single-molecule junctions.\cite{koch2006current,Aradhya_2013_Nat.Nanotechnol._p399,Zang_2020_NanoLett._p673,huang2019electric,PhysRevLett.123.246601,avriller2021photon,chen2021electron} To study chemical reactions in such scenarios, we have recently proposed a fully quantum mechanical simulation scheme, based on the hierarchical equations of motion (HEOM) approach. \cite{Erpenbeck_2019_J.Chem.Phys._p191101,Erpenbeck_2020_Phys.Rev.B_p195421,ke2021unraveling} Applications to current-induced bond rupture at metal surfaces, a process which is closely related to the breakdown of molecular functionalities,\cite{Sabater_2015_BeilsteinJ.Nanotechnol._p2338,Li_2015_J.Am.Chem.Soc._p5028,Li_2016_J.Am.Chem.Soc._p16159,Capozzi_NanoLett._2016_p3949--3954} showed that the HEOM method provides an accurate and efficient description of this process at high bias voltages.\cite{Erpenbeck_2020_Phys.Rev.B_p195421,ke2021unraveling} However, in the low bias voltage regime, reactions occur on a much longer timescale. Instead of tracking the dynamics of the slow transformation process from the reactants to the products, in such cases, the formulation of a reaction rate theory and the development of an efficient approach to calculate such rates are highly desirable.

To address this issue, in this work, we extend the flux correlation function formalism for the calculation of reaction rates to nonequilibrium scenarios. Within this extended formalism, the evaluation of an extended “Boltzmannized”  flux operator is required, which involves not only the molecular degrees of freedom (DoFs) but also the manifold of all electronic states of the metal surface/electrodes. To obtain this operator and calculate reaction rate in an efficient and exact way, we further present a mixed real and imaginary time HEOM approach for open quantum systems coupled to fermionic reservoirs. It is worth mentioning that the imaginary time HEOM approach can also be used as an effective tool for evaluating thermal equilibrium properties as well as preparing a correlated initial state for the real time HEOM method.

As a specific example to illustrate the validity and applicability of the proposed nonequilibrium flux correlation function formalism using the mixed real and imaginary time HEOM method, we consider intramolecular proton transfer reactions in a single-molecule junction. Proton transfer has been found to be involved in a variety of chemical and biological processes, \cite{Hammes-Schiffer_J.Phys.Chem.B_2008_p14108} and in this case quantum effects play an important role, which limits the applicability of approaches that are based on a classical description of the reactive nuclear DoFs.\cite{Tully_J.Chem.Phys._1971_p562,Shenvi_J.Chem.Phys._2009_p174107,Fischer_J.Chem.Phys._2011_p144102,Head_J.Chem.Phys._1995_p10137,Dou_J.Chem.Phys._2016_p24116,Galperin_J.Phys.Chem.Lett_2015_p4898,Lu_Prog.Surf.Sci._2019_p21,Lu_Phys.Rev.B_2012_p245444,Dzhioev_2011_J.Chem.Phys._p74701,Dzhioev_2013_J.Chem.Phys._p134103,Pozner_2014_NanoLett._p6244,Preston_Phys.Rev.B_2020_p155415,Erpenbeck_2018_Phys.Rev.B_p235452}  In the context of molecular electronics, it has been shown that current-induced proton transfer processes can be used to realize molecular nanoswitches and molecular transistors.\cite{Liljeroth_Science_2007_p1203--1206,Sobolewski_Phys.Chem.Chem.Phys._2008_p1243--1247,Benesch_J.Phys.Chem.C_2009_p10315--10318,Hofmeister_J.Mol.Model._2014_p1--5,Weckbecker_J.Phys.Chem.Lett._2020_p413--417} To control and design such devices requires a thorough theoretical study of current-induced proton transfer processes at molecule-metal interfaces, which was so far limited to perturbative master equations.\cite{Hofmeister_2017_J.Chem.Phys._p92317}

The remainder of this work is arranged as follows: 
In \Sec{model_method}, we introduce the nonequilibrium flux correlation function formalism for the specific example of current-induced reactions in molecular junctions. Furthermore, the description of the quantum dynamics with a correlated initial state and the real and imaginary time HEOM approach are outlined. Applications of the method to simulate current-induced proton transfer in molecular junctions are presented in \Sec{results}. In particular, we investigate the influence of the applied bias voltage and the molecule-lead coupling on the proton transfer rate. \Sec{conclusion} concludes with a summary.

\section{Model and Method}\label{model_method}
\subsection{\label{model}Model}
 As a specific example of a nonequilibrium scenario, we consider a molecular junction, where a single molecule is bounded to macroscopic electrodes. The full system is modeled by the Hamiltonian,
\begin{equation}
\label{total_Hamiltonian}
    H=H_{s}+H_{b}+H_{c},
\end{equation}
where $H_{s}$ corresponds to the molecular system,  $H_{b}$ to electrodes, and  $H_{c}$ describes their coupling.

The molecule comprises electrons and nuclei and the corresponding Hamiltonian is given by
\begin{equation}
H_{s}=T_n+H_{el},
\end{equation}
where $T_n=\sum_i p_i^2/2m_i$ represents the kinetic energy operator of the nuclei.
The variables $p_i$, $x_i$, and $m_i$ denote the momentum, position, and the effective mass of the $i$-th nuclear DoF, respectively. $H_{el}$ is the electronic Hamiltonian which includes electronic kinetic energy and the interactions among electrons and nuclei.

The electrodes serve as electron reservoirs and are represented as a continuum of non-interacting electrons, 
\begin{equation}
\label{Hamiltonian_reservoir}
H_{b}=\sum_{\alpha}\sum_{k}\epsilon_{\alpha k} c_{\alpha k}^{\dagger}c_{\alpha k},
\end{equation}
where $c_{\alpha k}^{\dagger}$ and $c_{\alpha k}$ are the creation and annihilation operators for an electron in the $k$-th state of electrode $\alpha$ with the corresponding energy 
\begin{equation}
  \epsilon_{\alpha k}=\epsilon_{\alpha k}^{(0)}+\mu_{\alpha}.  
\end{equation}
Here, $\mu_{\alpha}$ denotes the electrochemical potential of electrode $\alpha$, which depends on the applied bias voltage $\Phi$, and $\epsilon_{\alpha k}^{(0)}$ is the corresponding energy at zero bias voltage.

The coupling between the molecule and electrodes, which generally depends on the nuclear position $\mathbf{x}$, leads to electron transport. We assume that the interaction Hamiltonian takes the general form
\begin{equation}
\label{Hamiltonian_coupling}
H_{c}=\sum_{\alpha k }v_{\alpha k}s(\mathbf{x})c_{\alpha k}^{\dagger}d+v^*_{\alpha k}s(\mathbf{x})d^{\dagger}c_{\alpha k},
\end{equation}
where $v_{\alpha k}$ specifies the maximal coupling strength between the molecular electronic level and the $k$-th state in electrode $\alpha$. Here, a dimensionless real function $s(\mathbf{x})$ is introduced to model the dependence of the coupling on the nuclear position.
$d^{\dagger}$ $(d)$ denotes the operator for creating (annihilating) an electron in a specific molecular electronic level. Note that for notational simplicity, we consider a system with a single electronic level, but the extension to a multilevel system is straightforward.

\subsection{\label{flux_correlation_function_formalism}Nonequilibrium flux correlation function formalism}
The connection between the equilibrium reaction rate and the flux correlation function is well established.\cite{Miller_J.Chem.Phys._1983_p4889--4898,Thompson_J.Chem.Phys._1997_p142--150,Wang_J.Chem.Phys._2006_p174502,Craig_J.Chem.Phys._2007_p144503} However, reaction dynamics for molecular systems in contact with macroscopic electrodes is generally subject to nonequilibrium situations. For instance, the electrodes are usually held at different chemical potentials. As a consequence, the molecular system is driven out of equilibrium by the electron current. Our goal is to extend the equilibrium flux correlation function formalism and formulate a nonequilibrium quantum reaction rate theory. 
In the present work, we focus on a non-equilibrium scenario where the molecule is bound to two electrodes under an applied bias voltage $e\Phi=\mu_L-\mu_R$, and both leads have the same temperature $T$ (or the inverse temperature $\beta=1/k_{\rm B} T$, with $k_{\rm B}$ being the Boltzmann constant). 

First, we briefly review the basic ideas of the flux correlation function formalism pertaining to its application in this work. In the study of a unimolecular chemical reaction, the system is usually divided into a reactant and a product region, which are separated by a dividing surface located at $x_{\rm ds}$. The quantum-mechanical projection operator is given by
\begin{equation}
    h=\theta(x-x_{\rm ds}),
\end{equation}
where $\theta(x)$ denotes the Heaviside function, and its Heisenberg time-derivative yields the flux operator 
\begin{equation}
\label{flux_operator}
    F=\frac{i}{\hbar} [H,h].
\end{equation}

We assume that the system is initially located in the reactant region and represented by a non-stationary density operator $\rho_r$. Thereafter, the time evolution of the reactant and product population are given by
\begin{equation}
P_{r}(t)=\textrm{tr}\{(1-h)\rho(t)\}
=\textrm{tr}\{(1-h)e^{-\frac{i}{\hbar}Ht}\rho_re^{\frac{i}{\hbar}Ht}\}
\end{equation}
and
\begin{equation}
P_{p}(t)=1-P_{r}(t)=\textrm{tr}\{h\rho(t)\}
=\textrm{tr}\{he^{-\frac{i}{\hbar}Ht}\rho_re^{\frac{i}{\hbar} Ht}\},
\end{equation}
where $ \rho(t)=e^{-\frac{i}{\hbar}Ht}\rho_re^{\frac{i}{\hbar} Ht}$ denotes the density operator at time $t$. 

The definition of a rate constant for the transition from reactant to product (and vice versa) is based on the assumption that the population transfer is governed by first order kinetics at sufficiently long times,
\begin{equation}
\label{rate_process}
\frac{d}{dt}P_{r}(t)=-\frac{d}{dt}P_{ p}(t)=-kP_{r}(t)+k_{\rm b}P_{p}(t),
\end{equation}
in which $k$ and $k_{\rm b}$ are the forward and backward reaction rate constant, respectively.  
The forward rate constant $k$ can be formally expressed as\cite{Craig_J.Chem.Phys._2007_p144503}
\begin{eqnarray}
\label{rate_expression}
k &\equiv &\lim_{t\rightarrow \infty}k(t) \\
&=&\lim_{t\rightarrow \infty}\frac{C_{\rm f}(t)}{P_{ r}(0)+\chi[P_{r}(0)-1]-(1+\chi)\int_0^tC_{\rm f}(\tau) d\tau} \nonumber.
\end{eqnarray}
Here, $P_r(0)=\textrm{tr}\left\{ (1-h) \rho_r\right\}$ denotes the initial reactant population, and $\chi =k_{\rm b}/ k$ is the ratio of the backward and forward reaction rate, which depends on the temperature $\beta$ and applied bias voltage $\Phi$. It can be obtained as the ratio of the steady state solutions of $P_r(t)$ and $P_p(t)$,
\begin{equation}
\label{chi}
    \chi(\beta, \Phi) =\frac{P^{\rm ss}_{r}}{P^{\rm ss}_{p}},
\end{equation}
since the time derivative of the reactant (or product) population, as given in \Eq{rate_process}, is zero at steady state. Besides, we have introduced the reactive flux correlation function
\begin{equation}
\label{original_flux_side_correlation_function}
    C_{\rm f}(t)=-\dot{P}_{r}(t)=\textrm{tr}\left\{ 
    F_{\beta}e^{\frac{i}{\hbar}Ht} h e^{-\frac{i}{\hbar}Ht}
    \right\},
\end{equation}
and the “Boltzmannized” flux operator\cite{Miller_J.Chem.Phys._1983_p4889--4898}
\begin{equation}
\label{boltzmannized_flux_operatore}
  F_{\beta}=-\frac{i}{\hbar}[H,\rho_r].
\end{equation}

An alternative approach for calculating the rate constant is based on an assumption that the population transfer between reactants and products is much slower than any other dynamical processes in the system and its surroundings. When this assumption holds, the forward rate, e.g., can be calculated directly at \enquote{shorter times}, 
\begin{equation}
\label{approximate_rate}
k\approx \lim_{t\rightarrow t_{\rm plateau}}C_{\rm f}(t),
\end{equation}
where $t_{\rm plateau}$ is the time at which $C_{\rm f}(t)$ reaches a plateau. The above approximation is reliable when two conditions are met. The first condition is that the initial reactant population $P_r(0)$ is equal or very close to 1. Second, the reaction timescale is much longer than the timescales of other transient dynamical processes. When these two conditions are fulfilled, the loss of reactant population is negligible and the denominator in \Eq{rate_expression} is essentially unity over the timescale considered.  As such, one can avoid the calculation of $\chi$ and obtain the reaction rate by computing the flux correlation function using quantum dynamics approaches. 

Notice that the selected form of $\rho_r$ determines the explicit expression of the Boltzmannized flux operator in \Eq{boltzmannized_flux_operatore} and influences significantly the short-time dynamics of $C_{\rm f}(t)$. Nevertheless, different choices of $\rho_r$ which represent reliably the initial thermal ensemble (either "reactants" or "products") are expected to yield similar values for the rate constants. An optimal choice of $\rho_r$ should therefore be both a reliable representation of the initial ensemble and facilitate rapid convergence of $C_{\rm f}(t)$ to a plateau.
As illustrated in the supporting material, in the equilibrium cases ($\mu_L=\mu_R=0$), the flux correlation function shows the least oscillating behavior with the choice
\begin{equation}
\label{rho_r_eq}
    \rho^{\rm eq}_r=\frac{1}{Q_r^{\rm eq}}e^{-\beta H/2 }(1-h)e^{-\beta H/2 },
\end{equation}
corresponding to the Boltzmannized flux operator 
\begin{equation}
\label{equilibrium_F}
    F^{\rm eq}_{\beta}=\frac{1}{Q_r^{\rm eq}}e^{-\beta H/2 }Fe^{-\beta H/2 }.
\end{equation}

In nonequilibrium cases, a related choice is 
\begin{equation}
\label{initial_ss}
    \rho_r^{\rm ss}=\frac{1}{{Q^{\rm ss}_{r}}}e^{-\beta (H-Y)/2 }(1-h) e^{-\beta (H-Y)/2 }.
\end{equation}
Here, $Q_r^{\rm ss}=\mathrm{tr} \{(1-h)e^{-\beta (H-Y)}\}$, and the operator $Y$ is given formally as 
\begin{equation}
    Y=\sum_{\alpha=L/R}\Omega \mu_{\alpha}  N_{\alpha} \Omega^{-1}, 
\end{equation}
 where $\Omega=\lim_{t\rightarrow \infty} e^{\frac{i}{\hbar}Ht} e^{-\frac{i}{\hbar}{H_b t}}$ is the M\o ller operator, and $N_{\alpha}=\sum_{k}c_{\alpha k}^{\dagger}c_{\alpha k}$ represents the occupation number operator of the electrons in lead $\alpha$. 
The above expressions are built upon the nonequilibrium steady state (NESS) theory,\cite{Ness_Phys.Rev.E_2014_p_62119,Ness_Entropy_2017_p158} which shows that the nonequilibrium steady state can be described by a density matrix in a Gibbs-like form,
\begin{equation}
    \rho^{\rm ss}=e^{-\beta (H-Y) }/\mathrm{tr}\{e^{-\beta (H-Y) }\}.
\end{equation}
Using the fact that the operator $e^{-\beta (H-Y)}$ commutes with the Hamiltonian $H$, the Boltzmannized flux operator is expressed as
 \begin{equation}
 \label{steadystate_F}
  F^{\rm ss}_{\beta}=e^{-\beta (H-Y)/2 } F e^{-\beta (H-Y)/2 }/Q_r^{\rm ss},
\end{equation}
which reduces for vanishing bias voltage ($\mu_L=\mu_R=0$) to the equilibrium expression \Eq{equilibrium_F}. Expression \Eq{steadystate_F} for the flux operator, together with Eqs. (\ref{rate_expression}) - (\ref{original_flux_side_correlation_function}), generalizes the flux correlation function formalism to nonequilibrium situations and represent the main theoretical result of this work. 

Although Eqs. (\ref{initial_ss}) - (\ref{steadystate_F}) have an appealing structure, it is generally difficult to obtain the exact expression of $Y$. The expression simplifies significantly by replacing $H-Y$ by the zero-bias Hamiltonian $H-\sum_{\alpha}\mu_{\alpha} N_{\alpha}$, which corresponds to
\begin{equation}
\label{initial_state}
\rho_r^0=\frac{1}{Q^0_r}e^{-\frac{\beta}{2}(H-\sum_{\alpha}\mu_{\alpha}N_{\alpha})} (1-h) e^{-\frac{\beta}{2}(H-\sum_{\alpha}\mu_{\alpha}N_{\alpha})}
\end{equation}
with the partition function of the reactant
\begin{equation}
    Q^0_{r}=\mathrm{tr}\left\{(1-h)e^{-\beta (H-\sum_{\alpha}\mu_{\alpha}N_{\alpha})}\right\}. 
\end{equation}
The expression of $\rho_r^0$ in \Eq{initial_state} reduces also to the equilibrium expression, \Eq{rho_r_eq} for vanishing bias voltage. Although it does not lead to an intuitive form of a Boltzmannized flux operator, it is significantly easier to calculate than \Eq{initial_ss}. Furthermore, from  the physics perspective, notice that this choice of $\rho_r$ implies that the rate constant, as defined by \Eq{rate_expression} or (\ref{approximate_rate}), corresponds to the reaction induced by a sudden switch on of the bias voltage. Therefore, this rate should be directly comparable to experiments in which the reaction is induced by the applied voltage, and the reaction rate is much slower than the rate of the voltage ramp. Moreover, this approach applies also to voltage-induced dissociation rate calculations\cite{fung2019breaking,ke2021unraveling} by \Eq{rate_expression} or (\ref{approximate_rate}).

By performing the cyclic permutation within the trace expression, the reactive flux correlation function in \Eq{original_flux_side_correlation_function} is reformulated into the form
\begin{equation}
\label{nonequilibrium_flux_side_correlation_function_A}
    C_{\rm f}(t)=\mathrm{tr}\left\{ 
   Fe^{-\frac{i}{\hbar}Ht} \rho_r e^{\frac{i}{\hbar}Ht} 
    \right\}=
    \mathrm{tr}_s\left\{ 
   F\rho_r^s(t)
    \right\},
\end{equation}
which is valid for both choices, \Eq{initial_ss} and (\ref{initial_state}).
Here, we have used fact that the flux operator $F$ acts only on the system subspace, because the coupling Hamiltonian $H_c$ is only a function of $\mathbf{x}$ and commutes with the projection operator $h$.  The above equation indicates that the reactive flux correlation function  $C_{\rm f}(t)$ can be calculated as the expectation value of the flux operator with respect to the reduced density operator 
\begin{equation}
\label{reduced_rho}
 \rho_r^s(t) =\mathrm{tr}_b\{e^{-\frac{i}{\hbar}Ht} \rho_r e^{\frac{i}{\hbar}Ht} \}.   
\end{equation}
Above, $\mathrm{tr}_s$ and $\mathrm{tr}_b$ denote tracing over system and environmental DoFs, respectively. 

The numerically exact calculations of $\rho_r^s(t)$ can be performed with the HEOM approach. However, the method typically starts with a factorized initial state between the system and environment parts, which is not the case for $\rho_r$ in \Eq{initial_state}. 
Therefore, an additional imaginary time propagation is required.
Below we present a mixed real and imaginary time HEOM approach that can be used to efficiently obtain nonequilibrium flux correlation functions for chemical reaction dynamics in molecular systems that are in contact with fermionic reservoirs. 

\subsection{\label{method}Hierarchical equations of motion}
In the seminal work by Tanimura and Kubo,\cite{Tanimura_1989_J.Phys.Soc.Jpn._p101} the HEOM approach was proposed to investigate relaxation dynamics. Over the last three decades, considerable progress has been made to extend its applicability.\cite{Tanimura_2020_J.Chem.Phys._p20901,Yan_Chem.Phys.Lett._2004_p216--221,Xu_2005_J.Chem.Phys._p41103,Ishizaki_2005_J.Phys.Soc.Jpn._p3131,Shi_2009_J.Chem.Phys._p84105,Hu_2010_J.Chem.Phys._p101106,Kreisbeck_J.Chem.TheoryComput._2011_p2166--2174,Suess_Phys.Rev.Lett._2014_p150403,Hou_2015_J.Chem.Phys._p104112,Tang_2015_J.Chem.Phys._p224112,Tsuchimoto_J.Chem.TheoryComput._2015_p3859--3865,Ke_J.Chem.Phys._2016_p024101,Hsieh_2018_J.Chem.Phys._p14103,Nakamura_Phys.Rev.A_2018_p012109,Shi_J.Chem.Phys._2018_p174102,Erpenbeck_2018_J.Chem.Phys._p64106,Zhang_2020_J.Chem.Phys._p64107,Ikeda_J.Chem.Phys._2020_p204101} Tanimura developed an imaginary time HEOM for open systems coupled to bosonic environments to facilitate the calculations of thermal quantities.\cite{Tanimura_2014_J.Chem.Phys._p44114} Yan and coworkers significantly extend the method for studying charge transport in quantum impurity models.\cite{Jin_2008_J.Chem.Phys._p234703,Zheng_2008_J.Chem.Phys._p184112,Zheng_2013_Phys.Rev.Lett._p86601,Yan_2014_J.Chem.Phys._p54105,Ye_2016_WIREsComputMolSci_p608,Zhang_2017_J.Chem.Phys._p44105,Han_2018_J.Chem.Phys._p234108} Schinabeck \etal further extended it to investigate charge transport in nanosystems with electronic-vibrational couplings.\cite{Schinabeck_2016_Phys.Rev.B_p201407,Schinabeck_2018_Phys.Rev.B_p235429,Erpenbeck_2019_TheEuropeanPhysicalJournalSpecialTopics_p1981,Schinabeck_Phys.Rev.B_2020_p075422} Erpenbeck \etal formulated and applied the method for studying reaction dynamics at metal surfaces using real time propagation.\cite{Erpenbeck_2019_J.Chem.Phys._p191101} Here, we present a mixed real and imaginary time HEOM approach for open quantum systems which are initially correlated with fermionic environments.
The method is formulated within the reduced system framework, where the reduced density operator $\rho_r^s(t)$ is obtained by tracing over all reservoir DoFs. For the Hamiltonian given in \Sec{model}, this can be performed analytically in the Feynman path-integral formalism,\cite{Feynman__2010_p,Zinn-Justin__2010_p} and the expression of $\rho_r^s(t)$ reads

\begin{widetext}
\begin{eqnarray}
\label{density_matrix_path_integral_representation}
\rho^{s}_r(\varphi^+_t ,\tilde{\varphi}^-_t,t)&=
&\langle \varphi_t|\mathrm{tr}_b\left\{e^{-\frac{i}{\hbar}Ht}\hat{\rho}_re^{\frac{i}{\hbar}Ht}\right\}|\tilde{\varphi}_t\rangle\\
 &=& \nonumber
\frac{1}{Q_{r}}\int \mathrm{d} \varphi^+_0\mathrm{d}\varphi^-_0
\int \mathrm{d} \tilde{\varphi}^+_0\mathrm{d}\tilde{\varphi}^-_0
\nonumber \int_{\varphi^-_0}^{ \varphi^+_t} 
\mathcal{D}\varphi^+(t) \mathcal{D}\bm{\varphi}^-(t) 
 \int_{\varphi^+_0}^{ \tilde{\varphi}^-_0}
\mathcal{D}\bm{\phi}^+(\tau) \mathcal{D}\bm{\phi}^-(\tau)
\nonumber \int_{ \tilde{\varphi}^+_0 }^{ \tilde{\varphi}^-_t  } 
\mathcal{D} \tilde{\bm{\varphi}}^+(t) \mathcal{D}\tilde{\bm{\varphi}}^-(t) 
\\
&& \nonumber (1-h(\phi^+(\beta\hbar/2), \phi^-(\beta\hbar/2))) e^{\frac{i}{\hbar}S(\bm{\varphi}^+(t),\bm{\varphi}^-(t))
-\frac{1}{\hbar}S_{\beta}(\bm{\phi}^+(\tau),\bm{\phi}^-(\tau))
-\frac{i}{\hbar}S^*(\tilde{\bm{\varphi}}^+(t),\tilde{\bm{\varphi}}(t))} 
\nonumber \mathcal{F}.
\end{eqnarray}

Here, we have introduced the fermionic coherent state $\left| \varphi\right\rangle $, which is the eigenstate of $d$ with eigenvalue $\varphi^-$, i.e., $d\left| \varphi\right\rangle =\varphi^-\left| \varphi\right\rangle $. Its conjugation yields $\left\langle  \varphi\right| d^{\dagger} =\left\langle  \varphi\right| \varphi^+$ for the creation operator $d^{\dagger}$. $\varphi^+$ and $\varphi^-$ are independent Grassmann variables and satisfy the anticommutation relations $\{\varphi^{\sigma},\varphi^{\bar{\sigma}}\}=1$ and $\{\varphi^{\sigma}, \varphi^{\sigma}\}=0$. For conciseness, we introduce the symbol $\sigma=\pm$ and $\bar{\sigma}=-\sigma$.
$\bm{\varphi}^{\sigma}(t)$ and  $\tilde{\bm{\varphi}}^{\sigma}(t)$ are the forward and backward system paths along the real time axis $t$.
$\bm{\phi}^{\sigma}(\tau)$ denote the paths in the imaginary time $\tau$ within the range of $0$ and $\beta\hbar$. $S$ and $S_{\beta}$ denote the system action functional and their explicit expressions can be found in the supplementary material.

All the statistical information about the electron reservoirs and their influence on the system dynamics are encoded in the influence functional
\begin{equation}
\label{IF}
\begin{split}
\mathcal{F}=&e^{-\frac{1}{\hbar}\int_0^t\mathrm{d}s\int_0^{s}\mathrm{d}s'\sum_{\sigma=\pm}
\left[\varphi^{\bar{\sigma}}(s)+\tilde{\varphi}^{\bar{\sigma}}(s)\right]\sum_{\alpha}
\left[ G_{\alpha  }^{\sigma}(s-s') \varphi^{\sigma}(s')-G^{\bar{\sigma }*}_{\alpha }(s-s')\tilde{\varphi}^{\sigma}(s')  \right]}\\
&e^{\frac{i}{\hbar^2}\int_0^{t}\mathrm{d}s\sum_{\sigma=\pm}\left[\varphi^{\bar{\sigma}}(s)+\tilde{\varphi}^{\bar{\sigma}}(s)\right]\int_{0}^{\beta\hbar}\mathrm{d}\tau
\sum_{\alpha} X_{\alpha}^{\bar{\sigma} }(s+i\tau)\phi^{\sigma}(\tau)\mathrm{d}\tau +
\frac{1}{\hbar^2}\int_0^{\beta\hbar}\mathrm{d}\tau\int_{0}^{\tau}\mathrm{d}\tau'\sum_{\alpha\sigma}\phi^{\bar{\sigma}}(\tau)Y^{\sigma}_{\alpha }(\tau-\tau')\phi^{\sigma}(\tau')}.
\end{split}
\end{equation}

The thermal equilibrium correlation function of lead $\alpha$ is given by
\begin{equation}
\label{Fermi_correlation_function}
\begin{split}
G_{\alpha}^{\sigma}(t)
=\sum_{k} \frac{v_{\alpha k }^{*}v_{\alpha k} }{Q_{\alpha}}
\mathrm{tr}_{\alpha}\{  e^{\frac{i}{\hbar}H_{\alpha}t}c^{\sigma}_{\alpha k }e^{-\frac{i}{\hbar}H_{\alpha} t}c^{\bar{\sigma}}_{\alpha k }
e^{-\beta(H_{\alpha}-\mu_{\alpha}N_{\alpha})}
	\} 
=\frac{1}{2\pi}\int_{-\infty}^{\infty} e^{\frac{i \sigma}{\hbar}\epsilon t}\Gamma_{\alpha }(\epsilon)f_{\alpha}^{\sigma}(\epsilon)\mathrm{d}\epsilon,
\end{split}
\end{equation}
\end{widetext}
where $Q_{\alpha}=\mathrm{tr}_{\alpha}\left\{ e^{-\beta(H_{\alpha}-\mu_{\alpha}N_{\alpha})}\right\}$ is the partition function of lead $\alpha$. 
$\Gamma_{\alpha }(\epsilon)$ is the level-width function, which is defined as
\begin{equation}
\Gamma_{\alpha}(\epsilon)=2\pi \sum_{k}  v^*_{\alpha k} v_{\alpha k}\delta (\epsilon-\epsilon_{\alpha k}).
\end{equation}
In  what follows, a Lorentzian form is adopted,
\begin{equation}
\label{Lorentzian}
\Gamma_{\alpha}(\epsilon)=\frac{\Gamma_{\alpha} \Omega_{\alpha }^2}{\Omega_{\alpha }^2+(\epsilon-\mu_{\alpha})^2},
\end{equation}
where $\Gamma_{\alpha }$ quantifies the coupling strength between the molecular level and lead $\alpha$. $\Omega_{\alpha }$ is the bandwidth of the spectral energy distribution. The electron distribution in lead $\alpha$ at thermal equilibrium is given by the Fermi distribution function,
\begin{equation}
f_{\alpha}^{\sigma}(\epsilon)=\frac{1}{1+ e^{\sigma\beta (\epsilon- \mu_{\alpha})}}.
\end{equation}
Various sum-over-poles decomposition schemes for the Fermi distribution function have been proposed, including the Matsubara spectrum decomposition scheme, the Pad\'e spectrum decomposition scheme,
\cite{Hu_2010_J.Chem.Phys._p101106,Hu_2011_J.Chem.Phys._p244106}
and the Fano spectrum decomposition scheme.\cite{Cui_2019_J.Chem.Phys._p24110,Zhang_2020_J.Chem.Phys._p64107}  In this work, we employ the Pad\'e decomposition and expand the correlation functions in \Eq{Fermi_correlation_function} as a sum over exponential functions, 
\begin{equation}
\label{level_width_function_exponentials}
G_{\alpha}^{\sigma}(t-s) \simeq \sum_{p=0}^{P}\eta_{\alpha p}
e^{-\gamma^{\sigma}_{\alpha p}(t-s)},
\end{equation}
The explicit expressions of the coefficients $\{\eta_{\alpha p}\}$ and exponents $\{\gamma^{\sigma}_{\alpha p}\}$ are provided in the supporting material. 

The terms in the second line of \Eq{IF} describe the influence of the initial correlation between the system and reservoirs with the following correlation functions,
\begin{eqnarray}
\label{mixed_real_imag_time_correlation_function}
X_{\alpha}^{\bar{\sigma}}(t+i\tau)&=&
\frac{e^{-\bar{\sigma} \mu_{\alpha}\frac{\tau}{\hbar}}}{2\pi}\int_{-\infty}^{\infty}\mathrm{d}\epsilon e^{-\frac{i \bar{\sigma} }{\hbar} \epsilon(t+i\tau)}\Gamma_{\alpha}(\epsilon)
f^{\bar{\sigma}}_{\alpha}(\epsilon), \nonumber\\
&\simeq&
\sum_{p=0}^{P}\eta_{\alpha p}^*
e^{-\gamma^{\sigma}_{\alpha p}t}
e^{-i\gamma_{\alpha p}\tau},
\end{eqnarray}

and
\begin{widetext}
\begin{eqnarray}
\label{imag_time_correlation_function}
Y^{\sigma}_{\alpha}(\tau-\tau')
&=&\frac{1}{2\pi}\int_{-\infty}^{\infty}\mathrm{d}\epsilon
e^{\sigma\frac{ \epsilon-\mu_{\alpha} }{\hbar}(\tau-\tau') }\Gamma_{\alpha}(\epsilon)
f^{\sigma}_{\alpha}(\epsilon) \nonumber\\
&\simeq&\tilde{\eta}_{\alpha 0}\cos\Omega_{\alpha }\left(\frac{\beta}{2}-\frac{1}{\hbar}(\tau - \tau')\right) +\sum_{p=1}^{\infty}\tilde{\eta}_{\alpha p}
\left(\sin{\gamma_{\alpha p}(\tau-\tau')} \right).
\end{eqnarray}
with $\tilde{\eta}_{\alpha 0}=\eta_{\alpha 0}e^{i\beta\Omega_{\alpha}/2}$ and $\tilde{\eta}_{\alpha p}=i\eta_{\alpha p}$.

Given the above expansions of the correlations functions, we introduce a Grassmann variable,
\begin{eqnarray}
\label{B_grassman}
\mathcal{B}_{\bm{a}} &=&-\frac{i}{\hbar}\int_0^{t} e^{-\gamma^{\sigma}_{\alpha p}(t-s)}
 \left[ \eta_{\alpha p}\varphi^{\sigma}(s) 
 -\eta_{\alpha p}^{*}\varphi'^{\sigma}(s)  \right] \mathrm{d}s 
-\frac{\eta_{\alpha p}^*e^{-\gamma_{\alpha p}^{\sigma}t}}{\hbar}\int_0^{\beta\hbar}  \phi^{\sigma}(\tau) e^{-i\gamma_{\alpha p}\tau}\mathrm{d}\tau,
\end{eqnarray}
with a multi-index subscript  $\bm{a}=(\alpha,\sigma, p)$. For a given $\mathcal{B}_{\bm{a}}$, it can be understood as being related to the process where an electron from lead $\alpha$ is transferred to/back from ($\sigma=+/-$) the molecular electronic level with the characteristic correlation time $\gamma_{\alpha p}^{-1}$. 
To proceed, we introduce a set of auxiliary influence functionals $\mathcal{F}^{(n)}=\mathcal{B}_{\bm{a}_n}\cdots \mathcal{B}_{\bm{a}_1}\mathcal{F}$. By substituting the influence functional $\mathcal{F}$ in \Eq{density_matrix_path_integral_representation} with $\mathcal{F}^{(n)}$,
one readily constitutes a group of auxiliary density matrices (ADMs),  $\{\rho^{(n)}_{\bm{a}_1\cdots\bm{a}_n}(t)\}$, and their time-derivative yields the equation of motion\cite{Jin_2008_J.Chem.Phys._p234703}
\begin{equation}
\label{real_time_HQME}
\begin{split}
	\frac{\partial \rho^{(n)}_{\bm{a}_1,\cdots,\bm{a}_n}(t)}{\partial t}=&-\frac{i}{\hbar} \left[ H_{s},\rho^{(n)}_{\bm{a}_1,\cdots,\bm{a}_n}(t)\right] 
	-\left(\sum_{l=1}^{n} \gamma_{\bm{a}_l}\right)\rho^{(n)}_{\bm{a}_1,\cdots,\bm{a}_n}(t)\\
	&-\frac{i}{\hbar}\sum_{\bm{a}}
	\left(d^{\bar{\sigma}}s(\mathbf{x})
	 \rho^{(n+1)}_{\bm{a}_1,\cdots,\bm{a}_n,\bm{a}}(t) 
	  -(-1)^{n}\rho^{(n+1)}_{\bm{a}_1,\cdots,\bm{a}_n,\bm{a}}(t) d^{\bar{\sigma}}s(\mathbf{x})\right)\\
	 &
	-\frac{i}{\hbar} \sum_{l=1}^{n} (-1)^{n-l}\left(
	\eta_{\alpha_l p_l}d^{\sigma_l}s(\mathbf{x})\rho^{(n-1)}_{\bm{a}_1,\cdots,\bm{a}_{l-1},\bm{a}_{l+1},\cdots,\bm{a}_n}(t)
	+(-1)^{n}\eta_{\alpha_l p_l}^{*} \rho^{(n-1)}_{\bm{a}_1,\cdots,\bm{a}_{l-1},\bm{a}_{l+1},\cdots,\bm{a}_n}(t)d^{\sigma}s(\mathbf{x})\right).
\end{split}
\end{equation}
The above equation actually has the same form as the one obtained from a factorized initial condition.\cite{Jin_2008_J.Chem.Phys._p234703} However, the initial values of the ADMs are yet unkown, and they are given by definition as
\begin{equation}
\label{initial_values_for_real_time}
\begin{split}
	\rho^{(n)}_{\bm{a}_1,\cdots,\bm{a}_n}(\psi^+_0 ,\tilde{\psi}^-_0,t=0) =&\frac{1}{Q_{r}}
	\int^{\psi^+_0}_{ \tilde{\psi}^-_0}
	\mathcal{D}\bm{\phi}^+(\tau) \mathcal{D}\bm{\phi}^-(\tau)
	e^{-\frac{1}{\hbar}S_{\beta}(\bm{\phi}^+(\tau),\bm{\phi}^-(\tau))}\\
	&
	\exp\left\{\frac{1}{\hbar^2}\int_0^{\beta\hbar}\mathrm{d}\tau\int_{0}^{\tau}\mathrm{d}\tau'\sum_{\alpha\sigma} \phi^{\bar{\sigma}}(\tau)Y^{\sigma}_{\alpha }(\tau-\tau')\phi^{\sigma}(\tau') \right\}
			\left(1-h(\phi^+(\beta\hbar/2), \phi^-(\beta\hbar/2))\right)\mathcal{G}_{\bf{a}_n}
	\cdots\mathcal{G}_{\bf{a}_1}
\end{split}
\end{equation}
with
$\mathcal{G}_{\bf{a}}=-\frac{\eta_{\alpha p}^*}{\hbar}\int_0^{\beta\hbar}  \phi^{\sigma}(\tau) e^{-i\gamma_{\alpha p}\tau}\mathrm{d}\tau$ encoding the initial system-bath correlation information.

Following the ideas in Ref.\,\onlinecite{Tanimura_2014_J.Chem.Phys._p44114}, we introduce a set of imaginary time ADMs,
\begin{equation}
\label{imag_time_ADOs}
\begin{split}
	\tilde{\varrho}^{(n)}_{\bm{a}_1\kappa_1,\cdots,\bm{a}_n\kappa_n}(\psi^+_0 ,\tilde{\psi}^-_0,\tau,\lambda)=& 
	\frac{1}{\prod_{\alpha}Q_{\alpha}}\int^{\psi^+_0}_{ \tilde{\psi}^-_0}
	\mathcal{D}\bm{\phi}^+(\tau) \mathcal{D}\bm{\phi}^-(\tau)	e^{-\frac{1}{\hbar}S_{\tau}(\bm{\phi}^+(\tau),\bm{\phi}^-(\tau))} 	\left(1-\lambda	\theta(\tau-\beta\hbar/2)h(\phi^+(\beta\hbar/2), \phi^-(\beta\hbar/2))\right)\\
	&\exp\left\{\frac{1}{\hbar^2}\int_0^{\tau}\mathrm{d}\tau'\int_{0}^{\tau'}\mathrm{d}\tau''\sum_{\alpha\sigma} \phi_j^{\bar{\sigma}}(\tau')\tilde{\eta}_{\alpha 0}\cos\gamma_{\alpha 0}\left(\frac{\beta\hbar}{2}-\tau'+\tau''\right)
		 \phi^{\sigma}(\tau'') \right\} \\
	&\exp\left\{\frac{1}{\hbar^2}\int_0^{\tau}\mathrm{d}\tau'\int_{0}^{\tau'}\mathrm{d}\tau''\sum_{\alpha p\sigma} \phi^{\bar{\sigma}}(\tau')
	\tilde{\eta}_{\alpha p}
	\sin{\gamma_{\alpha p}(\tau'-\tau'')} 
	 \phi^{\sigma}(\tau'') \right\}
	 \tilde{\mathcal{G}}_{\mathbf{a}_n \kappa_n}\cdots\tilde{\mathcal{G}}_{\mathbf{a}_1 \kappa_1}.
\end{split}
\end{equation}
with $
\tilde{\mathcal{G}}_{\mathbf{a} \kappa}=\frac{1}{\hbar}\int_0^{\tau}  \phi^{\sigma}(\tau') \left[\kappa\cos(\gamma_{\alpha p}\tau')+\bar{\kappa}\sin(\gamma_{\alpha p}\tau')\right]\mathrm{d}\tau'$. The additional index $\kappa$ in the above definition takes only the value of either 0 or 1,  and $\bar{\kappa}=1-\kappa$. The imaginary time $\tau$ varies from 0 to $\beta\hbar$. Different from real time ADMs, all imaginary time ADMs are real-valued. 

By comparing  \Eq{initial_values_for_real_time} and \Eq{imag_time_ADOs}, one can verify that the real time ADMs at $t=0$ are related to the imaginary time ADMs of the same tier and can be obtained through the following transformation,
\begin{equation}
\label{transformation}
\rho^{(n)}_{\bm{a}_1,\cdots,\bm{a}_n}(t=0) =\frac{1}{Q_r^s}
\sum_{\kappa_1,\cdots,\kappa_n} \left(\prod_{l=1}^n \eta^*_{\alpha_l p_l}  \left(-\kappa_l+i\bar{\kappa}_l\right)\right)
	\tilde{\varrho}^{(n)}_{\bm{a}_1\kappa_1,\cdots,\bm{a}_n\kappa_n}(\beta\hbar,\lambda=1).
\end{equation}
The partition function of the reactant is given by 
\begin{equation}
Q_r^s=Q_{r}/\prod_{\alpha}Q_{\alpha}=\mathrm{tr}_s\left\{\tilde{\varrho}^{(0)}(\beta\hbar,\lambda=1)\right\}.
\end{equation}

Differentiating the imaginary time ADMs defined in \Eq{imag_time_ADOs} with respect to $\tau$, we have
\begin{equation}
\label{imaginary_time_HQME}
\begin{split}
	\frac{\partial \tilde{\varrho}^{(n)}_{\bm{a}_1\kappa_1,\cdots,\bm{a}_n\kappa_n}(\tau,\lambda)}{\partial \tau}=&
	-\frac{1}{\hbar}  H_{s}\tilde{\varrho}^{(n)}_{\bm{a}_1\kappa_1,\cdots,\bm{a}_n\kappa_n}(\tau,\lambda) \\
	&
	+\frac{1}{\hbar}\sum_{(\alpha \sigma \kappa)}d^{\bar{\sigma}}s(\mathbf{x})\tilde{\eta}_{\alpha 0}
	\left[\kappa\cos{\gamma_{\alpha 0}\left(\frac{\beta\hbar}{2}-\tau\right)} 
		- \bar{\kappa} \sin{\gamma_{\alpha 0}\left(\frac{\beta\hbar}{2}-\tau\right)}\right]
		\tilde{\varrho}^{(n+1)}_{\bm{a}_1\kappa_1,\cdots,\bm{a}_n\kappa_n,(\alpha \sigma 0 \kappa)}(\tau,\lambda) \\
	&	
	+\frac{1}{\hbar}\sum_{p=1}^P\sum_{\alpha \sigma\kappa}d^{\bar{\sigma}}s(\mathbf{x})\tilde{\eta}_{\alpha p}\left[\kappa \sin{\left(\gamma_{\alpha p}\tau\right)} 
	-\bar{\kappa}\cos{\left(\gamma_{\alpha p}\tau\right)}\right] 
	\tilde{\varrho}^{(n+1)}_{\bm{a}_1\kappa_1,\cdots,\bm{a}_n\kappa_n,(\alpha  \sigma p \kappa)}(\tau,\lambda)\\
	&+\frac{1}{\hbar} \sum_{l=1}^{n} (-1)^{n-l} 
	d^{\sigma_l}s(\mathbf{x}) 
	\left[\kappa_l\cos(\gamma_{\alpha_l p_l}\tau)+\bar{\kappa}_l\sin(\gamma_{\alpha_l p_l}\tau)\right]\tilde{\varrho}^{(n-1)}_{\bm{a}_1\kappa_1,\cdots,\bm{a}_{l-1}\kappa_{l-1},\bm{a}_{l+1}\kappa_{l+1},\cdots,\bm{a}_n\kappa_n}(\tau,\lambda),
\end{split}
\end{equation}
\end{widetext}
which is numerically solvable with the explicit initial condition. The zeroth tier imaginary time ADM is an identity matrix in the system subspace,  $\tilde{\varrho}^{(0)}(\tau=0)=\bm{I}$.
All other ADMs beyond the zeroth tier are zero, i.e., $\tilde{\varrho}^{(n>0)}(\tau=0)=\bm{0}$. 

It is worth noting that the hierarchies in \Eq{real_time_HQME} and \Eq{imaginary_time_HQME} both automatically terminate at a certain tier. The Grassmann variables $\mathcal{B}_{\bm{a}}$ and $\tilde{\mathcal{G}}_{\mathbf{a} \kappa}$ obey the anti-commutation relation, which implies that all indices in $\rho^{(n)}_{\bm{a}_1\cdots\bm{a}_n}(t)$ or $	\tilde{\varrho}^{(n)}_{\bm{a}_1\kappa_1,\cdots,\bm{a}_n\kappa_n}$ should be different.
The highest tier for nonzero real time ADMs is $L=2N_eN_{\alpha}(P+1)$, with $N_e$, $N_{\alpha}$, and $P$ being the number of molecular electronic levels, the number of reservoirs, and Pad\'e poles, respectively. For the nonzero imaginary time ADMs, the highest tier is increased to $2L$. In general, $L$ is very large, and a truncation of the hierarchy is necessary. To obtain the real time ADMs at $t=0$, as indicated by \Eq{transformation}, we should truncate the imaginary time hierarchy at least at the same tier as the real time hierarchy. However, in our simulations, we found that the imaginary time equation (\Eq{imaginary_time_HQME}) generally converges faster than the real time equation (\Eq{real_time_HQME}) with respect to the hierarchical truncation tier.  

With the zeroth tier real and imaginary time ADMs, one can obtain a variety of thermal and dynamical properties for the system of interest. For instance, the flux correlation function in \Eq{nonequilibrium_flux_side_correlation_function_A} is evaluated as the trace expression
\begin{equation}
  C_{\rm f}(t)=\mathrm{tr}_s\left\{F\rho^{(0)}(t)\right\}/Q_r^s, 
\end{equation}
and the system population in joint equilibrium with its reservoirs can be directly obtained as 
\begin{equation}
\label{equilibrium_population_x}
\rho_s^{\rm eq}(x)=\frac{\la x |\tilde{\varrho}^{(0)}(\beta\hbar,\lambda=0)|x\ra }{\mathrm{tr}_s\left\{\tilde{\varrho}^{(0)}(\beta\hbar,\lambda=0)\right\}}.
\end{equation}
Bath-related properties are enciphered in the higher-tier ADMs. For example, 
the electronic current flowing from lead $\alpha$ to the molecule can be defined as the time derivative of the average number of electrons in the lead, and is obtained through the first-tier ADMs,
\begin{equation}
\label{current}
\begin{split}
 	I_{\alpha}=&-e\frac{d\langle N_{\alpha} \rangle}{dt} \\
	=&\frac{ie}{\hbar}
\sum_{p} \mathrm{tr}_s \left\{ d s(\mathbf{x})\rho^{(1)}_{(\alpha,+,p)}- d^{\dagger}s(\mathbf{x}) \rho^{(1)}_{(\alpha,-,p)}
\right\},   
\end{split}
\end{equation}
where $e$ denotes the electron charge. 
The interaction energy at position $\mathbf{x}$ is also related to the first-tier ADMs,
\begin{equation}
\label{interaction_energy}
\begin{split}
\langle H_c(\mathbf{x})\rangle 
=&
\sum_{\alpha p} s(\mathbf{x})\la \mathbf{x}|\mathrm{tr}_{s_e} \left\{  d\rho^{(1)}_{(\alpha,+,p)}+d^{\dagger}\rho^{(1)}_{(\alpha,-,p)}
\right\}|\mathbf{x}\ra,
\end{split}
\end{equation}
where  $\mathrm{tr}_{s_e}$ denotes tracing over system electronic DoFs. 
The derivation details are provided in the supporting material.

\begin{figure}
	\begin{minipage}[c]{0.4\textwidth}		
			\raggedright a) \\
			\hspace{2cm}
	\includegraphics[width=\textwidth]{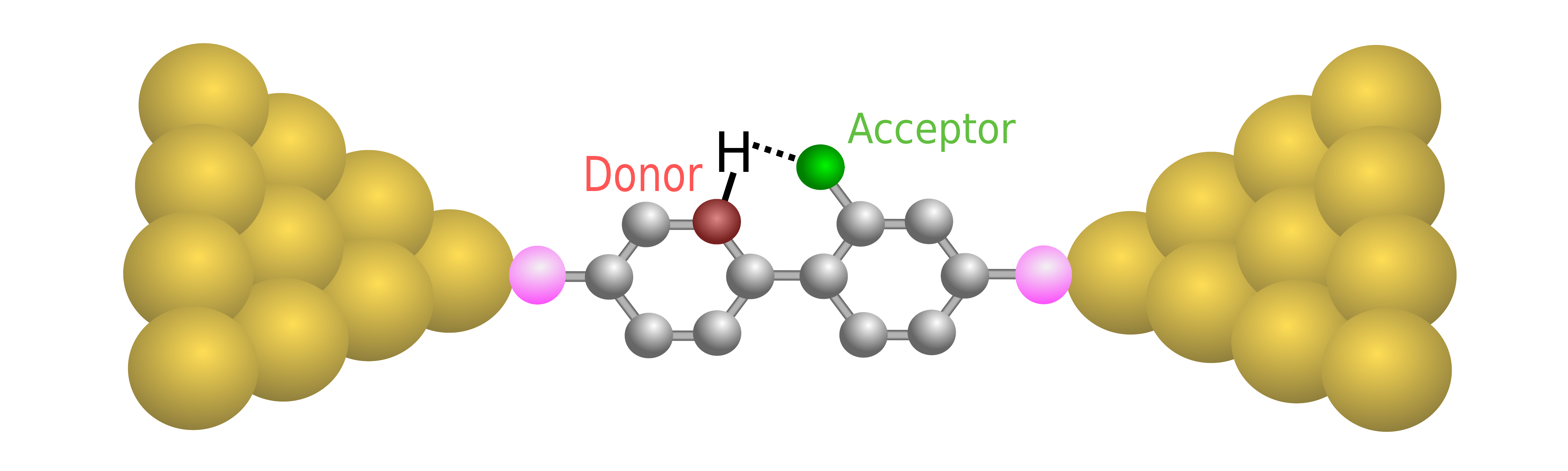}\\
	\end{minipage}
		\begin{minipage}[c]{0.4\textwidth}		
			\raggedright b) \\
	\includegraphics[width=\textwidth]{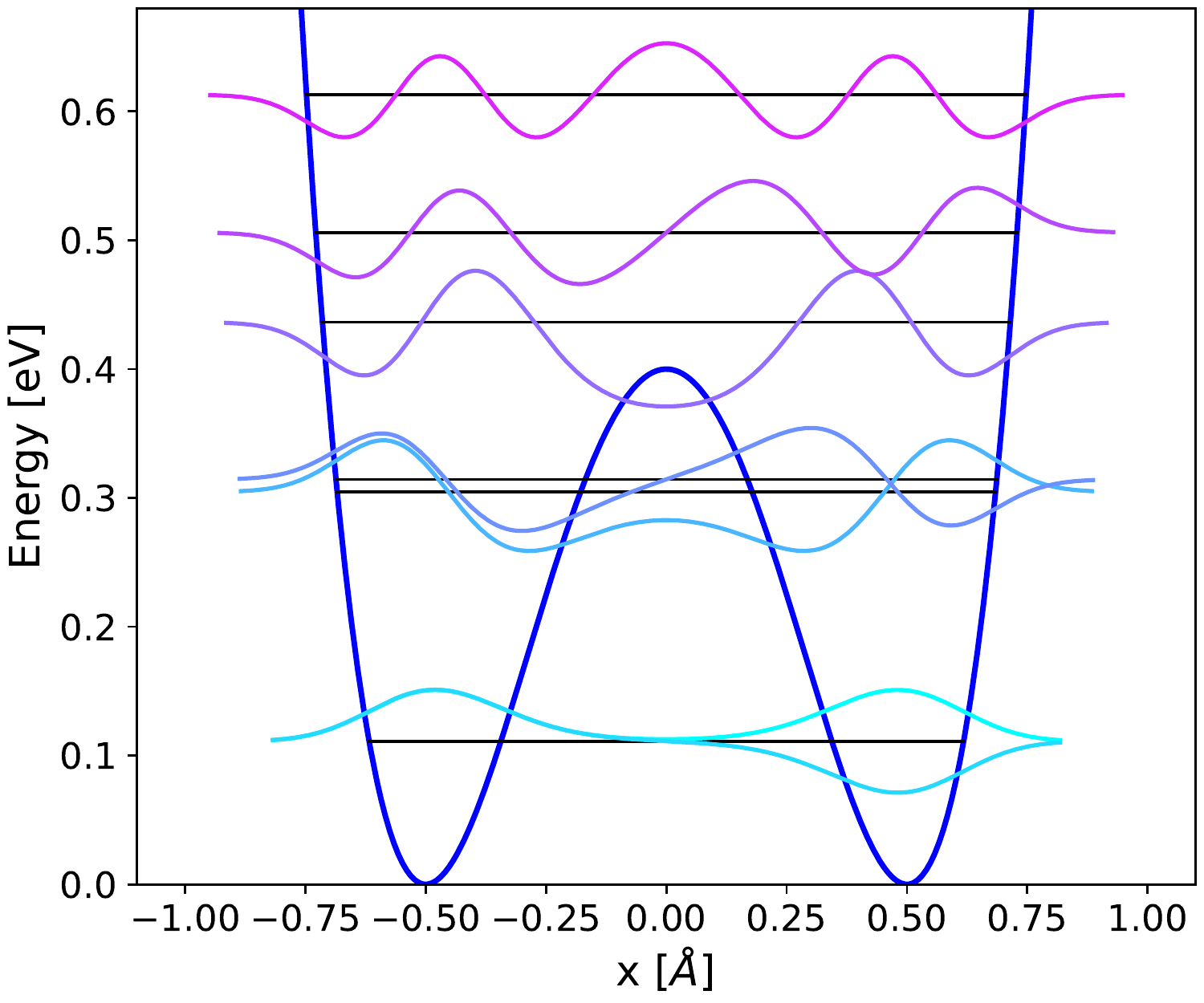}\\
	\end{minipage}
	\begin{minipage}[c]{0.4\textwidth} 		
			\raggedright c) \\
	\includegraphics[width=\textwidth]{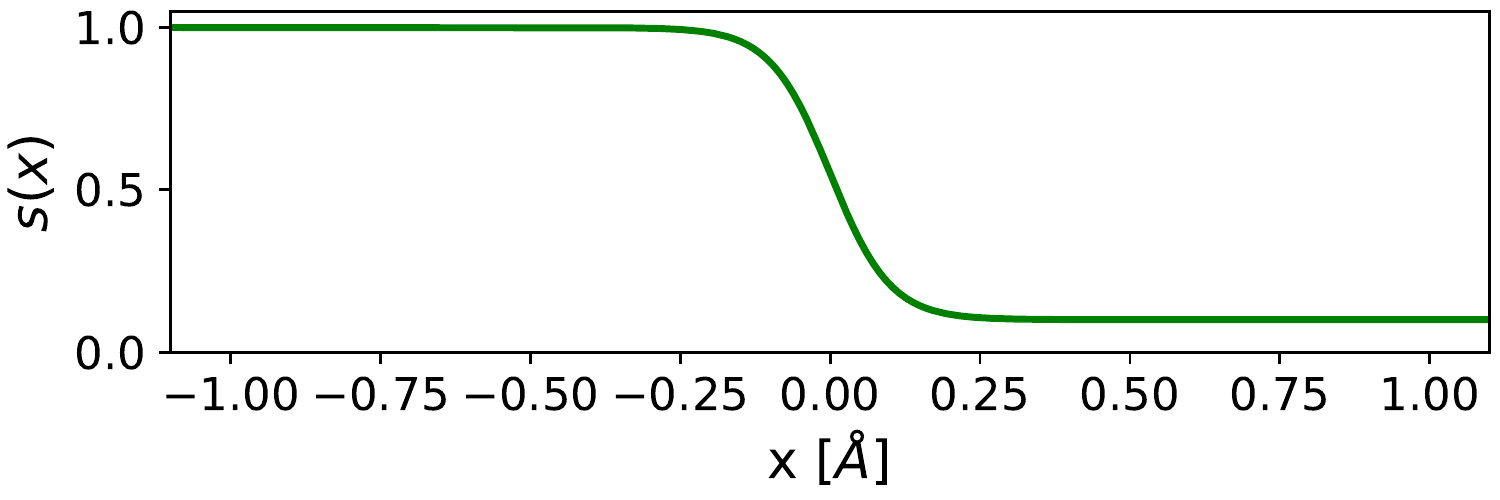}
	\end{minipage}
	\caption{(a) Schematic representation of intramolecular proton transfer in a molecular junction. (b) Double well potential for the proton transfer reaction model of the isolated molecule (see \Eq{V_0} with $l=0.5$ \AA \,and $V_b=0.4$ eV). The black horizontal and colored wavy lines denote the energy and wavefunction of the eigenstates, respectively. The first and the second state are nearly degenerate.
	(c) Switch function defined in \Eq{switch_function} with $\sigma=0.1$.	}
		\label{energy_profile}	
\end{figure}
\section{\label{results} Results}
In this section, we use the method outlined above to investigate the reaction rate of intramolecular proton transfer dynamics in a molecular junction. We first introduce the model system used and provide the simulations details. Then, results for a broad range of applied bias voltages and molecule-lead couplings are analyzed and discussed. 

\subsection{Model}
 We consider a one-dimensional model for studying the proton transfer dynamics in a molecular junction. The system Hamiltonian is given by
 \begin{equation}
H_{s}=\frac{p^2}{2m}+V_0(x) +\epsilon(x) d^{\dagger}d,
 \end{equation}
 As shown in \Fig{energy_profile} (a), the proton is translocated between the donor and acceptor moieties within the molecule. 
We shall assume that the proton transfer between donor and acceptor sites changes the molecule-lead couplings, but without loss of generality, the local potential energy surfaces at donor and acceptor sites are assumed to be identical. Therefore, the potential energy  of the neutral molecule along the reaction coordinate is assumed to have the form of a symmetric double-well
\begin{equation}
\label{V_0}
    V_0(x)=\frac{V_b}{l^4}(x+l)^2(x-l)^2,
\end{equation}
which is characterized by an energy barrier $V_b$ at $x=0$.
Two local minima are located at $x=\pm l$, corresponding to two stable conformations of an isolated molecule. 

The model was used in previous work to investigate the influence of proton transfer on the conductance in molecular junctions.\cite{Hofmeister_J.Chem.Phys._2017_p092317}  Specifically, \Fig{energy_profile} (b) depicts the potential with the parameters $l=0.5\textrm{ \AA}$ and $V_b=0.4$ eV. 
The energy levels and wavefunctions of the seven lowest eigenstates are shown as well.
The vibrational frequency is characterized as the curvature near the local minima, i.e., $\hbar\omega_0=\left.\sqrt{\frac{1}{m}\frac{\partial^2 V_0(x)}{\partial x^2}}\right|_{x=\pm l}=0.2$ eV (the proton mass is 1 amu), which characterizes the timescale of the nuclear motion. 
In general, the potential energy surface for the charged molecule may be different from the neutral one. Here, we assume for simplicity that the charging only leads to a constant energy shift by the electron affinity $\epsilon_0$, i.e., $\epsilon(x)=\epsilon_0$. In the following calculations, we adopt a value of $\epsilon_0=0.1$ eV. More general cases where $\epsilon(x)$ is a function of $x$ will be the subject of future work.

Besides, we assume that the dependence of molecule-lead coupling on the proton position (see \Eq{Hamiltonian_coupling}) is characterized by the function 
\begin{equation}
\label{switch_function}
    s(x)=\frac{1+\sigma}{2}-\frac{1-\sigma}{2}\tanh(x),
\end{equation}
which is visualized in \Fig{energy_profile} (c) with $\sigma=0.1$. As a result, in the current model, the conductance is larger when the proton is located in the left well and smaller in the right well. Furthermore, we point out that because of the assumed uniform charging energy at any proton position, the coupling of the nuclear motion to the electrons arises exclusively from the coordinate-dependent coupling to the metal electrons.

\subsection{Simulation details}
To obtain the proton transfer rate, a convenient choice of the diving surface is $x_{\rm ds}=0$, but we should stress that, in principle, the obtained rate is invariant to the choice of the dividing surface. For all calculations presented in this work,  we assume that the molecule is coupled symmetrically to the left and right leads, $\Gamma_{L/R}=\Gamma/2$ and a temperature of $T=300$ K is adopted.
The proton degree of freedom is represented using the sine discrete variable representation (DVR).\cite{Colbert_1992_J.Chem.Phys._p1982,Echave_1992_Chem.Phys.Lett._p225,Seideman_1992_J.Chem.Phys._p4412} A number of 26 DVR grid points is adequate to give converged results. The bandwidth in \Eq{Lorentzian} is taken as $\Omega_{L/R}=1$ eV. 
At least 15 Pad\'e poles are required. The hierarchical equations (\Eq{real_time_HQME} and \Eq{imaginary_time_HQME}) are truncated at the third tier ($L=3$) to assure convergence unless otherwise specified. We use a simple truncation scheme for both real and imaginary time equations, namely, setting the ADMs beyond the $L$th tier to zero. Both real and imaginary time equations are propagated using the fourth-order Runge-Kutta method. The ratio of reactant and product steady state population $\chi$ in  \Eq{rate_expression}  is obtained employing an efficient iterative steady state solver developed by Kaspar \etal\cite{Kaspar_J.Phys.Chem.A_2021_p5190}. 
The numerical integration is performed on NVIDIA Tesla V100 GPUs.

\subsection{Reaction at different bias voltage regimes}
First, we study the proton transfer dynamics in nonequilibrium situations, where the molecule is connected to two electrodes at a nonzero applied bias voltage. The bias voltage drops symmetrically on both electrodes, i.e., $\mu_L=-\mu_R=e\Phi/2$. The coupling to the two leads is fixed at $\Gamma_{L/R}=0.05$ eV.

\begin{figure}[H]
	\centering
	\begin{minipage}[c]{0.4\textwidth}		
		\raggedright a) $\Phi=$ 0.1 V\\
		\hspace*{0.2cm}
		\includegraphics[width=0.9\textwidth]{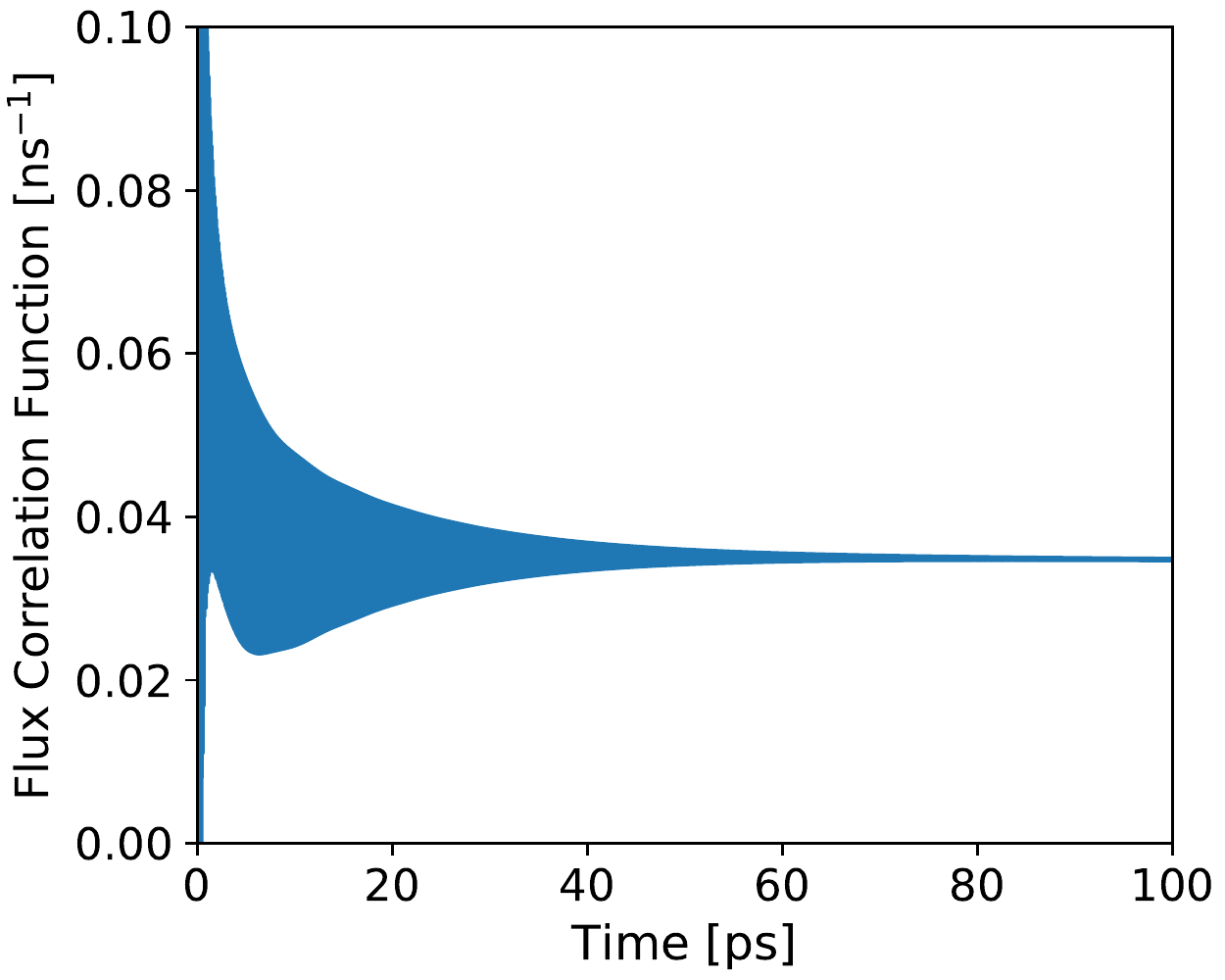}\\
	\end{minipage}
	\begin{minipage}[c]{0.4\textwidth} 		
		\raggedright b) $\Phi=$ 0.2 V\\
				\hspace*{0.2cm}
		\includegraphics[width=0.9\textwidth]{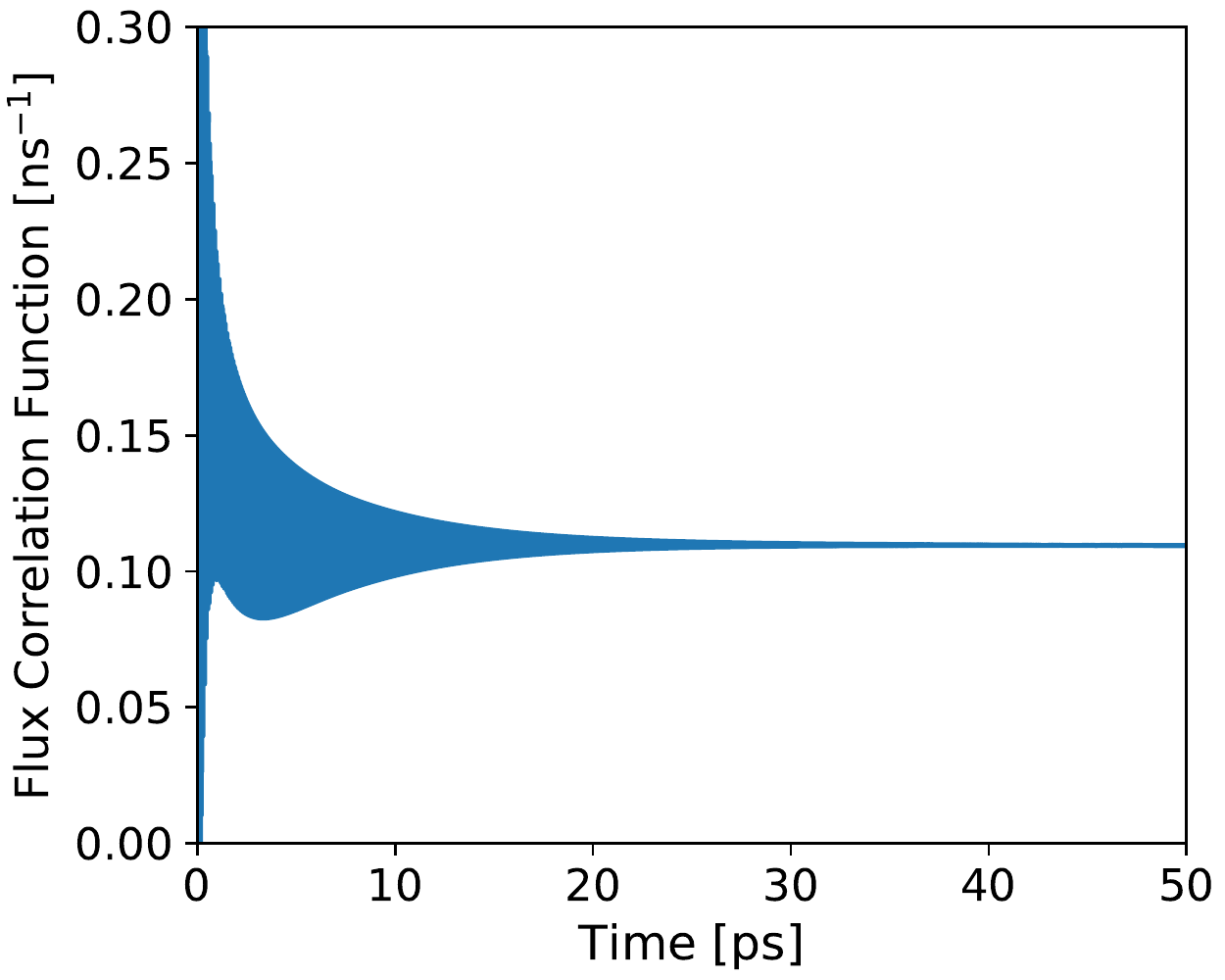}\\
	\end{minipage}
	\begin{minipage}[c]{0.4\textwidth} 		
		\raggedright c) $\Phi=$ 4 V\\
				\hspace*{0.2cm}
		\includegraphics[width=0.9\textwidth]{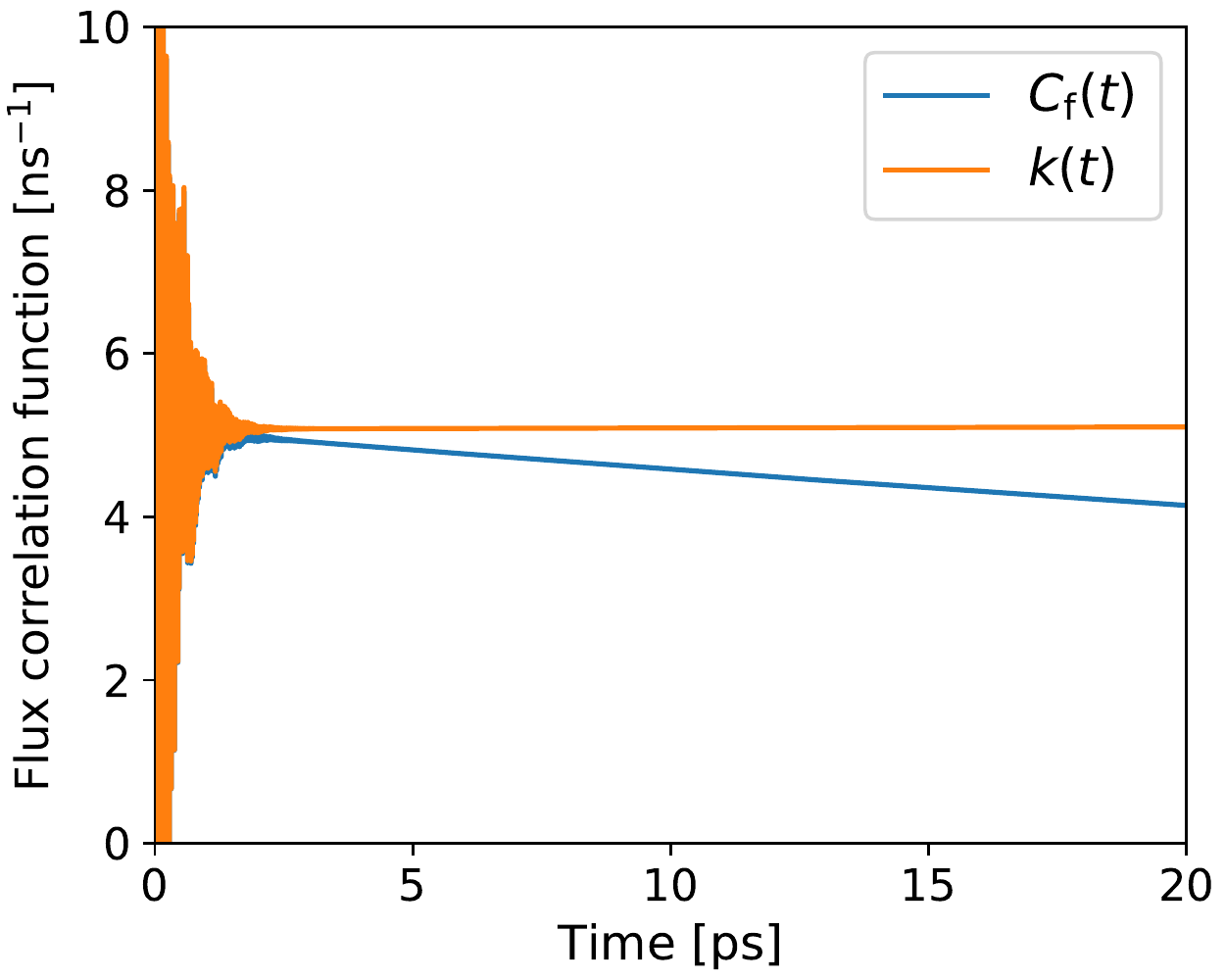}\\
	\end{minipage}
		\caption{Nonequilibrium flux correlation function $C_{\rm f}(t)$ (see \Eq{nonequilibrium_flux_side_correlation_function_A}) for three different bias voltages in the off-resonant, near resonant, and deep resonant transport regime, respectively. In addition, the  rate  expression $k(t)$ (see \Eq{rate_expression}) at a high bias voltage is shown in (c). In (a) and (b) for $\Phi= 0.1$ and 0.2 V, $k(t)$ is not shown as the results overlap  $C_{\rm f}(t)$. }
		\label{non-equilibrium_correlation_function}
\end{figure}

\Fig{non-equilibrium_correlation_function} (a) and (b) show
the flux correlation function  $C_{\rm f}(t)$ for two different bias voltages in the off and near-resonant transport regime, respectively. 
At low biases, for instance, $\Phi=0.1$ and $0.2$ V, the flux correlation functions exhibit significant oscillations with a period of about 21 fs ($\sim 0.2$ eV), which can be assigned to the vibrational motion at the bottom of the double well.  After tens of picoseconds, the oscillations are damped out and $C_{\rm f}(t)$ reaches a well-defined plateau. In these cases, the approximate expression \Eq{approximate_rate} can be used to obtain the nonequilibrium rate constant for current-induced proton transfer process, as the time separation condition is satisfied and the initial population of the reactant is very close to one ($P_r(0)=0.99993$).  

In the resonant transport regime, as illustrated in \Fig{non-equilibrium_correlation_function} (c) for $\Phi=4$ V, the oscillations of the flux correlation function vanish after a few picoseconds and the reaction takes place faster. As a consequence, the flux correlation function  $C_{\rm f}(t)$ decreases over time because the timescale separation is no longer satisfied and the reactant population changes significantly from its initial value. Nevertheless, the reaction obeys for longer time the first order kinetic scheme, \Eq{rate_process}, with the rate constant $k$ given by the generalized expression,  \Eq{rate_expression}, which exhibits a plateau, as shown by the orange line. 

\begin{figure}
	\centering
		\includegraphics[width=0.45\textwidth]{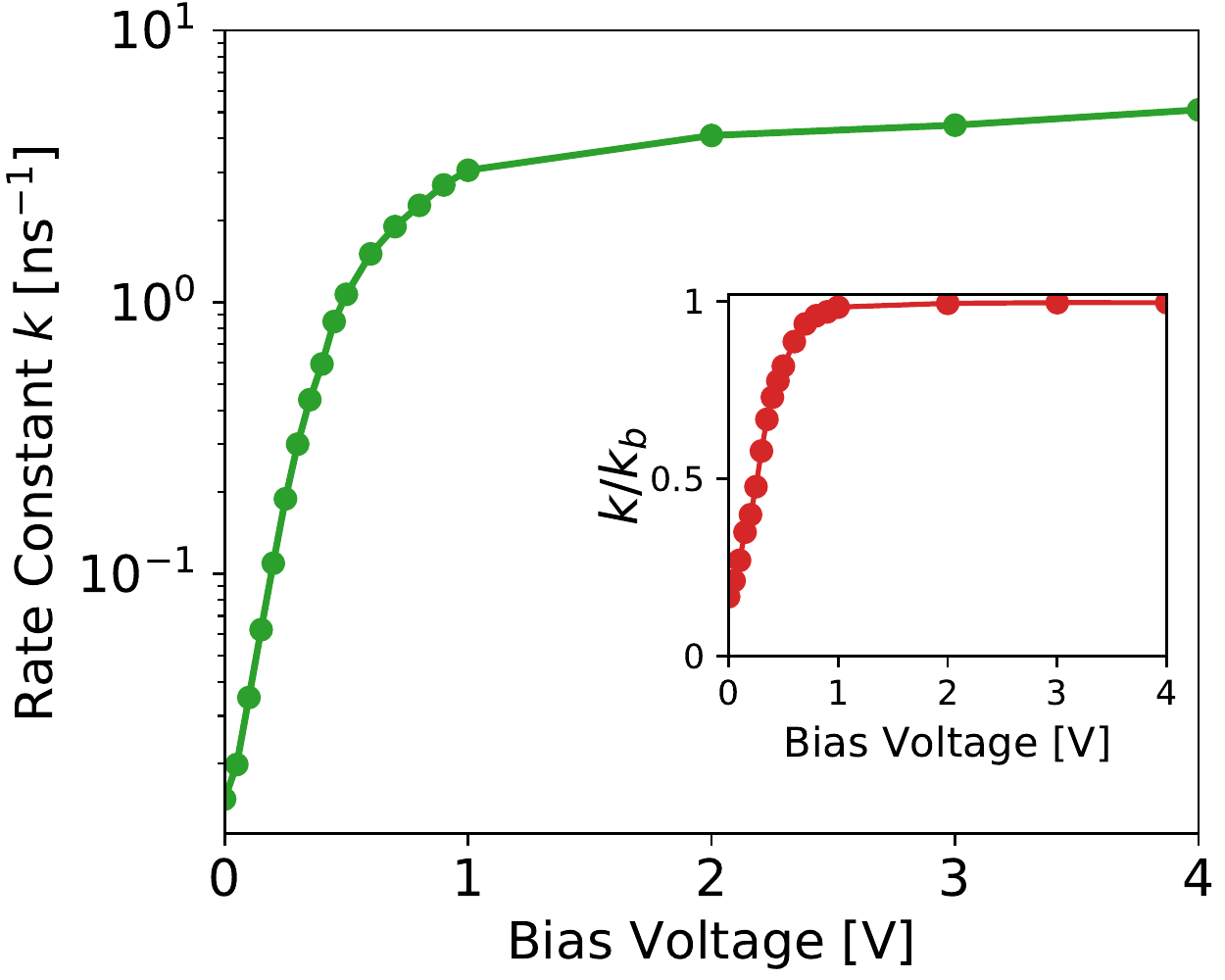}\\
		\caption{Forward reaction rate constant $k$ and the ratio $k/k_b$ 
		as a function of the applied bias voltage.}
		\label{chi_k_neq}
\end{figure}

\Fig{chi_k_neq} displays the forward reaction rate $k$ and the ratio $k/k_b$ in the inset against the bias voltages $\Phi$ ranging from 0 to 4 V.  The reaction rate increases with bias voltage in a bimodal way with a steep rise at lower bias voltages and a slow increase at higher bias voltages. 

At zero bias voltage, there is no electrical current flowing through the molecule. However, the proton transfer reaction rate is not zero due to quantum tunneling and thermal activation effects.  It should be noted that due to the small mass of the proton, the vibrational frequency $\omega_0$ is relatively high. As such, even at the room temperature, proton transfer due to quantum tunneling (rate  roughly estimated as $\omega_0\exp\left(- V_b/(\hbar\omega_0/2\pi)\right)$) dominates over thermally activation processes (rate $\sim \omega_0\exp\left(-V_b/(k_B T)\right)$). Interestingly, although we assume a symmetric molecular potential surface, the forward reaction rate is smaller than the backward reaction rate ($k/k_b=0.167$). This is due to the coupling of the molecule to the leads, which is a function of the proton position $x$, where more details will be elaborated in \Sec{influence_gamma}.

Turning on the bias voltage, the electrons transfer into and out of the molecule with a timescale characterized by $\hbar/\Gamma$, and this process can be accompanied by vibrational excitations. Thus, the proton transfer is additionally induced by current-induced vibrational excitation,\cite{Elste_Appl.Phys.A_2008_p345--354} which is much faster than quantum tunneling and thermal activation processes. At low bias voltages, more vibrationally coupled electron transport channels are opened up with increasing bias voltage. As a result, the forward reaction rate is significantly increased.  However, when the bias voltage is sufficiently large,
the energy transmitted into the molecule by a single electron transferring from the left lead to the right lead can overcome the reaction barrier for both forward and back reaction.
In this case, by further increasing the bias voltage, the reaction rate increases only slightly and the ratio $k/k_b$ approaches 1.

\subsection{\label{influence_gamma}Influence of molecule-lead coupling}
It is known that the coupling between molecule and leads can have a profound impact on the molecular  properties.\cite{Molen_J.Phys.:Condens.Matter_2010_p133001,Chen_Chem.Rev._2020_p2879} 
In this subsection, we investigate the influence of the molecule-lead coupling strength on the proton transfer rate and molecular properties.

\begin{figure}[h]
	\centering
	\begin{minipage}[c]{0.45\textwidth}		
		\raggedright a) $\Phi=0$ V\\
		\hspace*{0.2cm}
		\includegraphics[width=\textwidth]{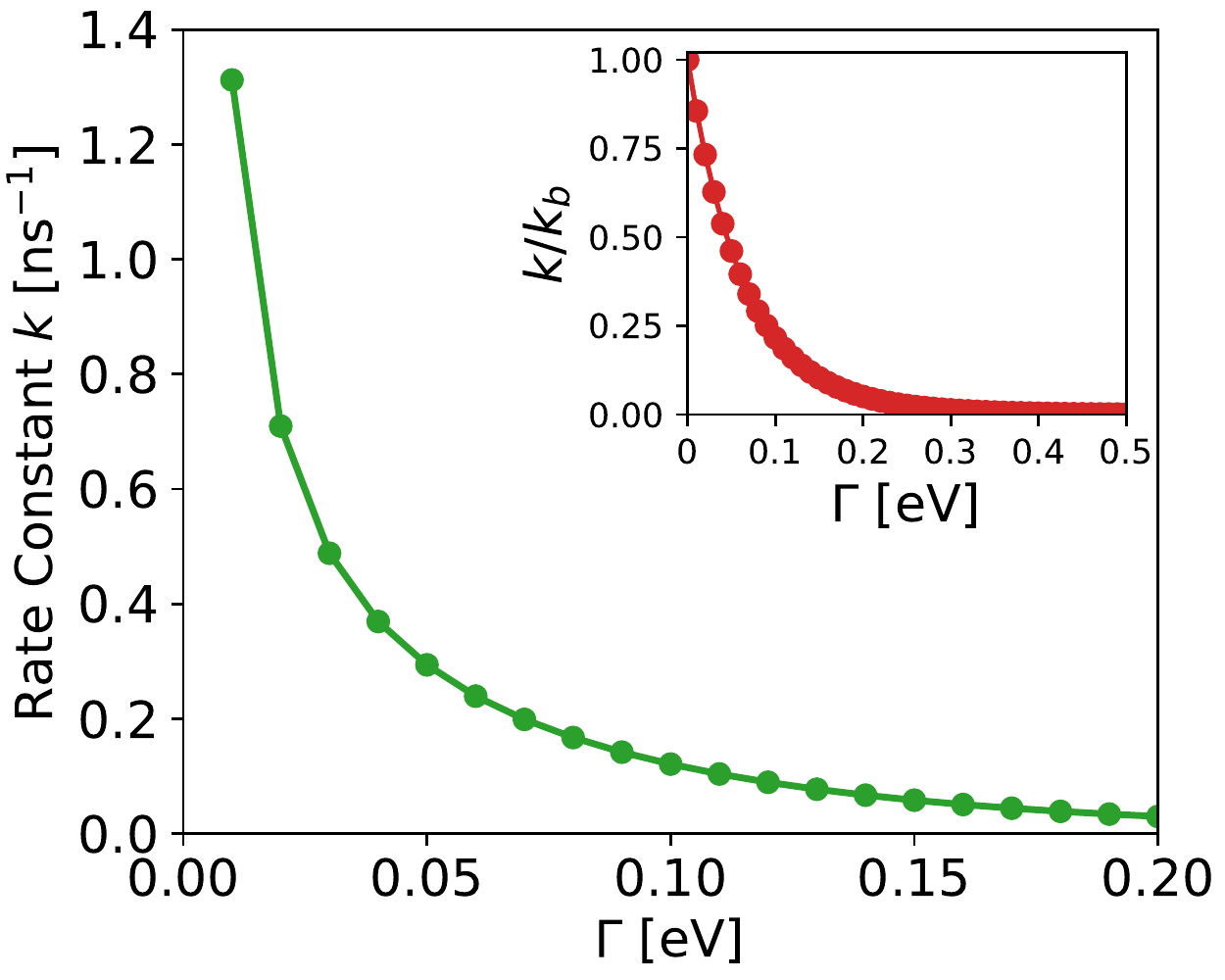}\\
	\end{minipage}
		\begin{minipage}[c]{0.45\textwidth}		
		\raggedright b) $\Phi=4$ V\\
		\hspace*{0.2cm}
		\includegraphics[width=\textwidth]{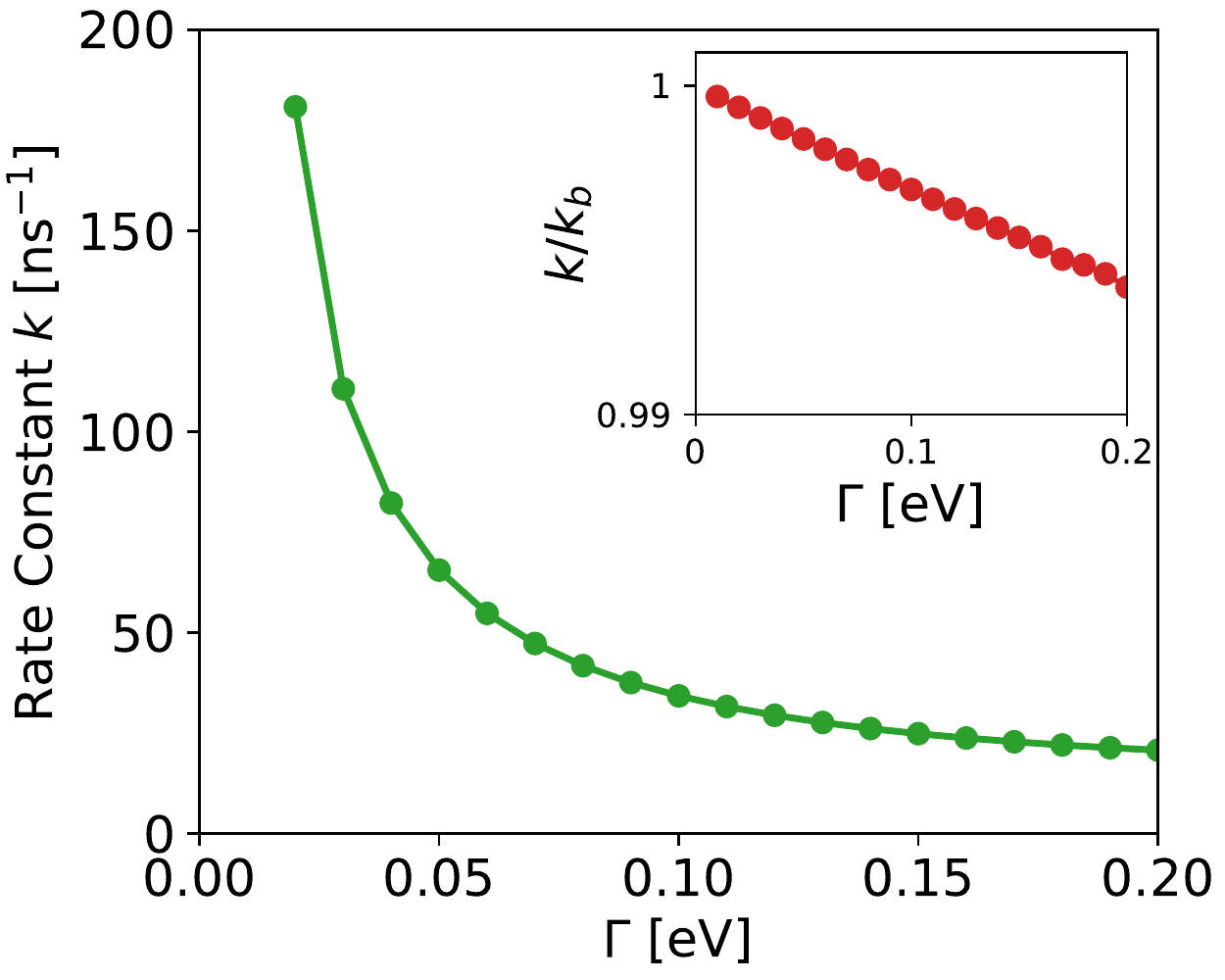}\\
	\end{minipage}
	\caption{Forward reaction rate constant $k$ as a function of the molecule-lead coupling strength $\Gamma$ for two different bias voltages. Shown in the inset are the ratios of forward and backward reaction rate $k/k_b$ against $\Gamma$. The barrier height is $V_b=0.3$ eV. The results for $\Phi=$ 0 V are obtained with a fourth-tier hierarchical truncation.}
	\label{k_eq}
\end{figure}

\Fig{k_eq} displays the proton transfer rate as a function of the molecule-lead coupling strength in the range of $\Gamma=0.01$ to 0.2 eV for both equilibrium ($\Phi=0$ V) and nonequilibrium ($\Phi=4$ V) cases. The ratios of the forward and back reaction rates $k/k_b$ are shown in the insets. We use a smaller barrier height $V_b=0.3$ eV, which facilitates the calculation of the reaction rate as the oscillations are damped at a shorter timescale. 

At equilibrium with zero bias voltage, we observe that the forward reaction rate $k$ and the ratio $k/k_b$ decrease monotonically with increasing molecule-lead coupling strength $\Gamma$. For $\Gamma>0.2$ eV, the flux correlation function still shows a strongly oscillatory structure on the scale of hundreds of picoseconds (data not shown), and the propagation to a much longer timescale is required to obtain the reaction rate. 

At finite bias voltage in the resonant transport regime, as shown in \Fig{k_eq} (b) for $\Phi=4$ V, the reaction rates are about two orders of magnitude larger, but exhibit the same dependence on $\Gamma$. This suggests that inherent molecular properties are the main reason for the behavior rather than the external bias voltage. 

\begin{figure}[H]
	\centering
		\begin{minipage}[c]{0.4\textwidth} 		
		\raggedright a)\\
		\includegraphics[width=0.9\textwidth]{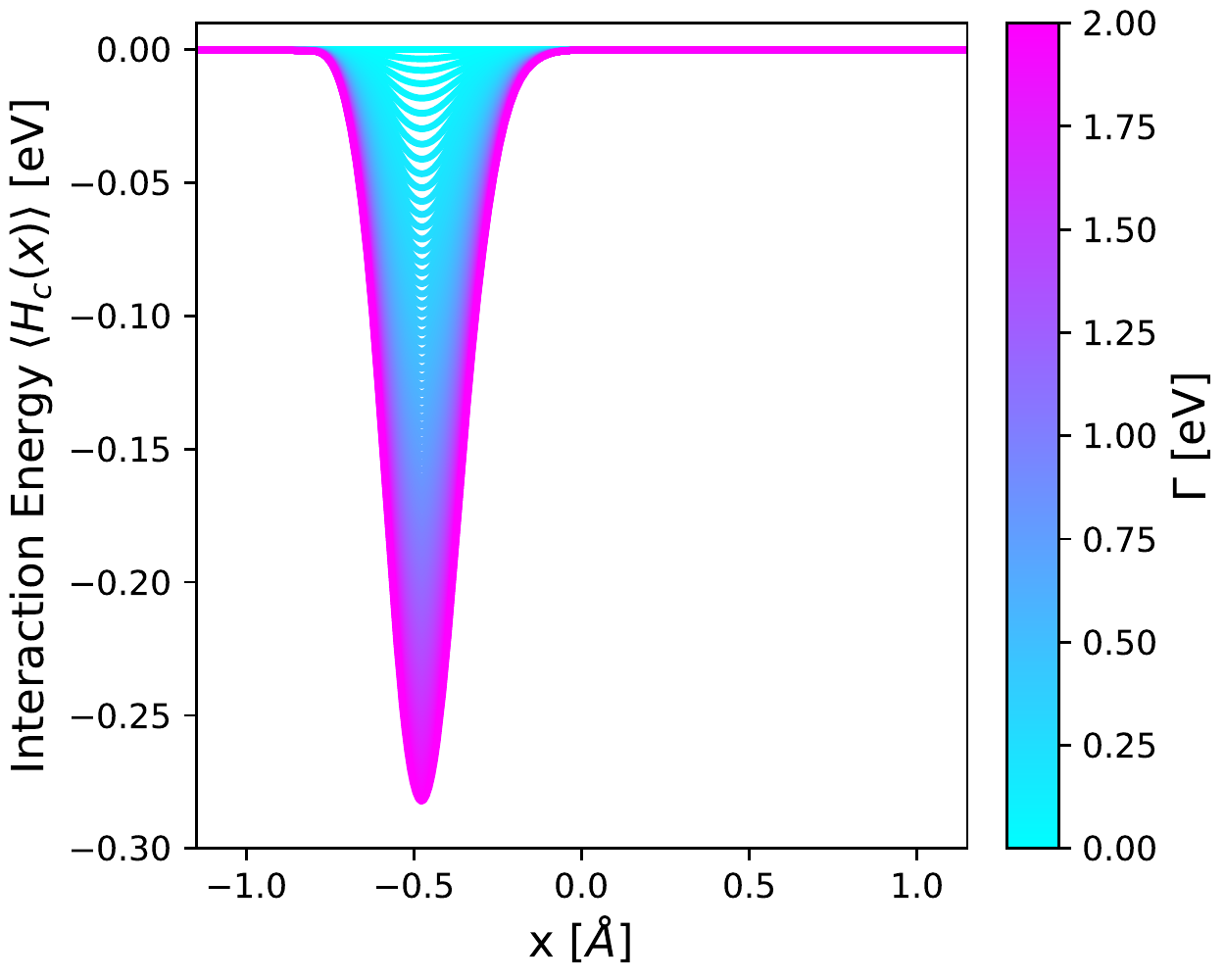}
		\begin{minipage}[c]{0.4\textwidth} 		
		\end{minipage}\\
		\raggedright b)\\
		\includegraphics[width=0.9\textwidth]{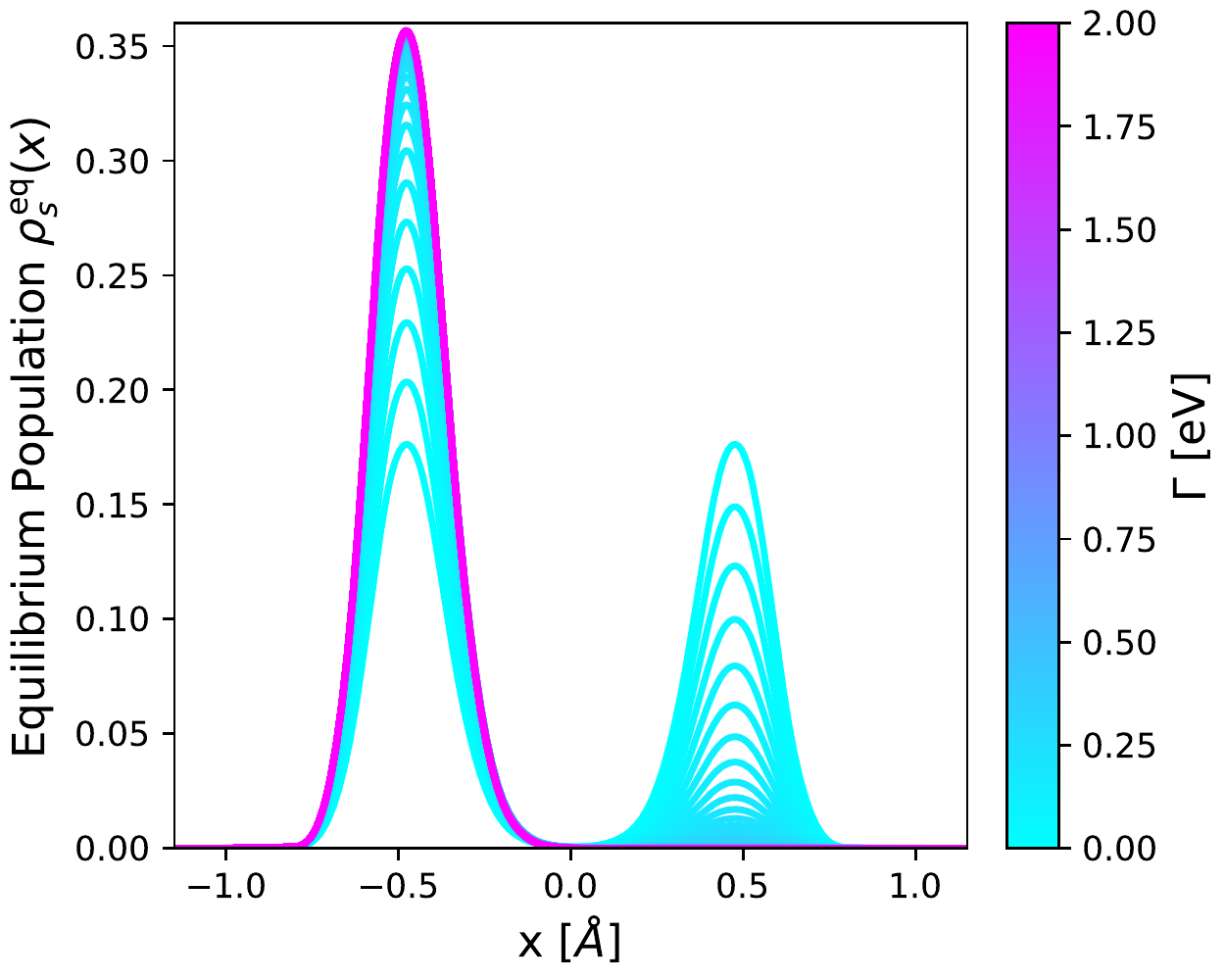}
	\end{minipage}\\
	\begin{minipage}[c]{0.4\textwidth}		
		\raggedright c) \\
		\includegraphics[width=0.9\textwidth]{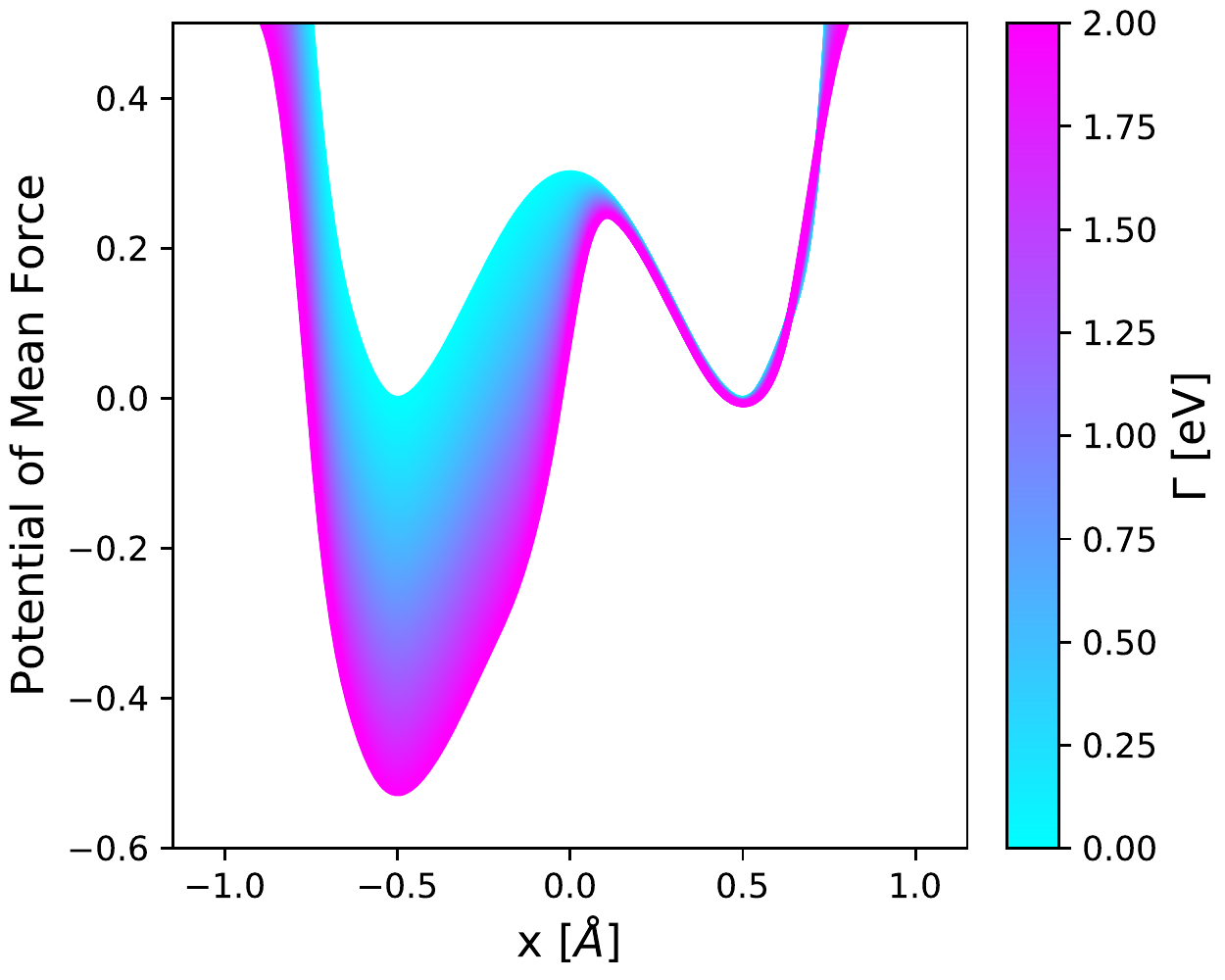}\\
	\end{minipage}
		\caption{Interaction energy between the system and reservoirs (a) and equilibrium populations (b) as well as the potential of mean force (c) at $300$ K. The definitions of the above quantities are given in \Eq{interaction_energy}, \Eq{equilibrium_population_x}, and \Eq{potential_of_meanforce},  respectively, and are obtained with a fourth-tier hierarchical truncation. The lines with color varying from cyan to purple correspond to different molecule-lead couplings ranging from 0 to 2 eV with a spacing of 0.02 eV. 
		Other parameters are: bias voltage $\Phi=0$ V, barrier height $V_b=0.3$ eV.}
		\label{equilibrium_population}
\end{figure}
To understand this behavior, we calculate the interaction energy $\la H_c(x)\ra$  (\Eq{interaction_energy}), the equilibrium population $\rho^{\rm eq}_s(x)$ (\Eq{equilibrium_population_x}),  as well as the potential of mean force $V_{\rm PMF}(x)$ for the reactive nuclear motion at equilibrium, which is given by\cite{Dou_J.Chem.Phys._2016_p74109}
\begin{equation}
\label{potential_of_meanforce}
\begin{split}
    V_{\rm PMF}(x)=&-\frac{1}{\beta} \ln 
    \la x|
    {\rm tr}_e \{e^{-\beta  H_{el}}\}
    |x\ra
  \end{split}
\end{equation}
where  $\mathrm{tr}_{e}$ denotes tracing over all electronic DOFs. ${\rm tr}_e \{e^{-\beta  H_{el}}\}$ can be obtained exactly through the imaginary time HEOM approach, as presented in \Eq{imaginary_time_HQME}. The results are displayed in \Fig{equilibrium_population} for various molecule-lead coupling strength ranging from 0 to 2 eV. 

For vanishing molecule-lead coupling, the interaction energy is zero and the potential of mean force reproduces the symmetric double-well potential of the isolated neutral molecule. The equilibrium populations in the reactant and product region are equal, thus we have $k/k_b=1$.   Increasing the molecule-lead coupling $\Gamma$, the interaction energy $\langle H_c(x)\rangle $ shows a dip in the reactant region with a minimum at around $x=-0.5$ \AA, which is due to the stronger coupling of the molecule to the leads when the proton is placed closer to the donor site.  The negative interaction energy in the reactant region leads to a stabilization effect.  As a consequence, the molecule is more likely to populate the reactant region for increasing $\Gamma$, and $k/k_b$ decreases. In particular, for strong coupling $\Gamma>0.5$ eV, the equilibrium population in the product region as well as the ratio $k/k_b$ approach zero, as shown in  \Fig{equilibrium_population} (b) and the inset of \Fig{k_eq} (a), respectively. 
The potential of mean force becomes asymmetric and the barrier top shifts towards the positive $x$ direction. The barrier height between the bottom of the left well and the barrier top is enhanced with increasing $\Gamma$, which results in a decrease of the forward reaction rate $k$. 

The results shown here indicate the crucial role of a coordinate-dependent molecule-lead coupling in reshaping the reactive potential surface and the reaction rates. This not only deepens our understanding of chemical reactions at molecule-metal interfaces. More importantly, it may be utilized to steer the chemical reaction in the desired direction by engineering the molecule-lead coupling strength.

\section{\label{conclusion}Conclusions}
We have presented a nonequilibrium reaction rate theory based on an extension of the flux correlation function formalism to nonequilibrium conditions. Moreover, to efficiently calculate flux correlation functions within this approach, we have proposed a mixed real and imaginary time HEOM method for open quantum systems that are coupled to fermionic reservoirs. The combined method can be used to calculate rate constants of chemical reactions in nonequilibrium situations in a fully quantum mechanical and numerically exact way.  

As an illustrative example, we have considered a model of intramolecular proton transfer in a molecular junction and calculated reaction rates for current-induced proton transfer for a range of different bias voltages and molecule-lead coupling strengths. For small bias voltages, time-scale separation allows to use directly the flux correlation function to determine the rate constant. For higher bias voltages, a renormalized flux-correlation function formulation is required to calculate the rate constant. The results also show that a strong molecule-lead interaction that depends on the reaction coordinate can reshape the effective potential energy surface and influence the reaction dynamics profoundly. 

In the present work, we have considered a simple model for proton transfer with a single reaction coordinate. Using the recent extension of the HEOM method, which combines it with the matrix product state representation,\cite{yaling_heom_tt} it should also be applicable to more general models and different reactions involving several nuclear degrees of freedom and multiple electronic states of the molecule. 

\section{Acknowledgements}
The authors thank Jakob B\"atge helpful discussions. This work was supported by the German Research Foundation (DFG). Y.K. was supported by the Alexander von Humboldt Foundation. 
The work of A.E. was funded by the DFG (grant 453644843). U.P. acknowledges support by the Israel Science foundation (1243/18) and by the U.S. - Israel Binational Science foundation. Furthermore, the authors acknowledge support by the state of Baden-Württemberg through bwHPC
and the German Research Foundation (DFG) through grant no INST 40/575-1 FUGG (JUSTUS 2 cluster).

\section*{Supplementary Material}
See the supplementary material for the  details  of  (1) the explicit expressions of the coefficients $\{\eta_{\alpha p}\}$ and exponents $\{\gamma^{\sigma}_{\alpha p}\}$ in \Eq{level_width_function_exponentials}; (2) derivation details of the interaction energy and current expression in \Eq{interaction_energy} and \Eq{current}, respectively; (3) the flux correlation function $C_{\rm f}(t)$ calculated with different initial density operators. 

\section*{Data Availability}
The data that support the findings of this study are available from the corresponding author upon reasonable
request.

\footnotesize

\end{document}